\documentclass[preprint,3p,times,twocolumn]{elsarticle}

\usepackage{amssymb}

\journal{Journal of Non-Newtonian Fluid Mechanics}

\usepackage{graphicx}
\usepackage{dcolumn}
\usepackage{bm}

\usepackage[T1]{fontenc}
\DeclareFontFamily{T1}{calligra}{}
\DeclareFontShape{T1}{calligra}{m}{n}{<->s*[1.44]callig15}{}
\DeclareMathAlphabet\mathcalligra   {T1}{calligra} {m} {n}
\DeclareMathAlphabet\mathzapf       {T1}{pzc} {mb} {it}
\usepackage[utf8]{inputenc}
\usepackage[T1]{fontenc}
\usepackage{mathptmx}
\usepackage{etoolbox}
\usepackage[T1]{fontenc}
\usepackage[final]{microtype}
\usepackage[x11names]{xcolor}
\usepackage{epstopdf}
\usepackage{mathrsfs}
\usepackage{amsmath} 
\usepackage{rsfso}
\usepackage{lineno}
\usepackage{hyperref}
\hypersetup{
	colorlinks,
	linkcolor={blue!85!black},
	citecolor={blue!85!black},
	urlcolor={blue!85!black},
	pdfauthor = {Vedad Dzanic, Christopher Soriano From, Emilie Sauret},
	}
 
\DeclareMathAlphabet{\mathsfit}{T1}{\sfdefault}{\mddefault}{\sldefault}
\SetMathAlphabet{\mathsfit}{bold}{T1}{\sfdefault}{\bfdefault}{\sldefault}
\providecommand\bnabla{\boldsymbol{\nabla}}
\providecommand\bcdot{\boldsymbol{\cdot}}
\DeclareMathAlphabet\mathsfbi            {OT1}{cmss}{m}{sl}
\IfFileExists{t1phv.fd}
{\DeclareTextFontCommand\textsfbi{\usefont{T1}{phv}{b}{it}}
	\DeclareMathAlphabet\mathsfbi            {T1}{phv}{b}{it}
}{}
\IfFileExists{ot1phv.fd}
{\DeclareTextFontCommand\textsfbi{\usefont{OT1}{phv}{b}{it}}
	\DeclareMathAlphabet\mathsfbi            {OT1}{phv}{b}{it}
}{}

\begin{document}

\begin{frontmatter}



\title{Lattice Boltzmann methods for simulating non-Newtonian fluids: A comprehensive review}


\author[inst1]{Vedad Dzanic\corref{cor1}}
\ead{v2.dzanic@qut.edu.au}
\fntext[label2]{}
\cortext[cor1]{}



\affiliation[inst1]{School of Mechanical, Medical, and Process Engineering, Queensland University of
Technology, Brisbane, 4001, QLD, Australia}

\author[inst1]{Qiuxiang Huang}

\author[inst2]{Christopher Soriano From}
\affiliation[inst2]{Department of Chemical Engineering, University of Manchester, Oxford Road, Manchester,
M13 9PL, UK.}
\author[inst1]{Emilie Sauret}

\begin{abstract}
Non-Newtonian fluids encompass a large family of fluids with additional nonlinear material properties, contributing to non-trivial flow behaviour that cannot be captured through a single constant viscosity term. Common non-Newtonian characteristics include shear-thinning, shear-thickening, viscoplasticity, and viscoelasticity, commonly encountered in everyday fluids, such as ketchup, blood, toothpaste, mud, etc., as well as practical applications involving porous media, cosmetics, food processing, and pharmaceuticals. Due to the complex nature of these fluids, accurate computational fluid dynamics simulations are essential for predicting their behaviour under various flow conditions. Recent advancements have highlighted the growing trend of using the lattice Boltzmann method to solve such complex flows, owing to its ability to handle intricate boundary conditions, ease of including additional multiphysics, and providing computationally efficient parallel simulations. Since the initial review over a decade ago [Phillips \& Roberts, \textit{IMA J. Appl. Math.} \textbf{76}, 790-816 (2011)], significant advancements have been made to the lattice Boltzmann method to simulate non-Newtonian fluids. Here, we present a comprehensive review of different lattice Boltzmann techniques used to solve non-Newtonian fluid systems, specifically dealing with shear-dependent viscosity, viscoplasticity, and viscoelasticity. In addition, we discuss various benchmark cases that validate these approaches and highlight their growing application to realistic and challenging complex flow problems. We further address outstanding issues in current lattice Boltzmann models, as well as future directions for numerical advancement and application.

\end{abstract}



\begin{keyword}
Lattice Boltzmann methods \sep Non-Newtonian fluids \sep Shear-thinning \sep Viscoelasticity \sep Yield stress
\end{keyword}

\end{frontmatter}


\section{Introduction}
\label{sec:Intro}
\noindent A fluid's ability to adhere to Newton's law of constant viscosity is far from a common generalisation but more so a rare exception. Most fluids realistically follow either a nonlinear relationship between the shear stress and shear strain, or can exhibit additional anomalous flow features, 
such as a combined fluid-solid or fluid-elastic behaviour. 
These complex types of fluids belong to the more broader class of fluids appropriately termed “non-Newtonian” fluids. 
\begin{figure*}[t]
	\centering
	\includegraphics[width=0.9\textwidth]{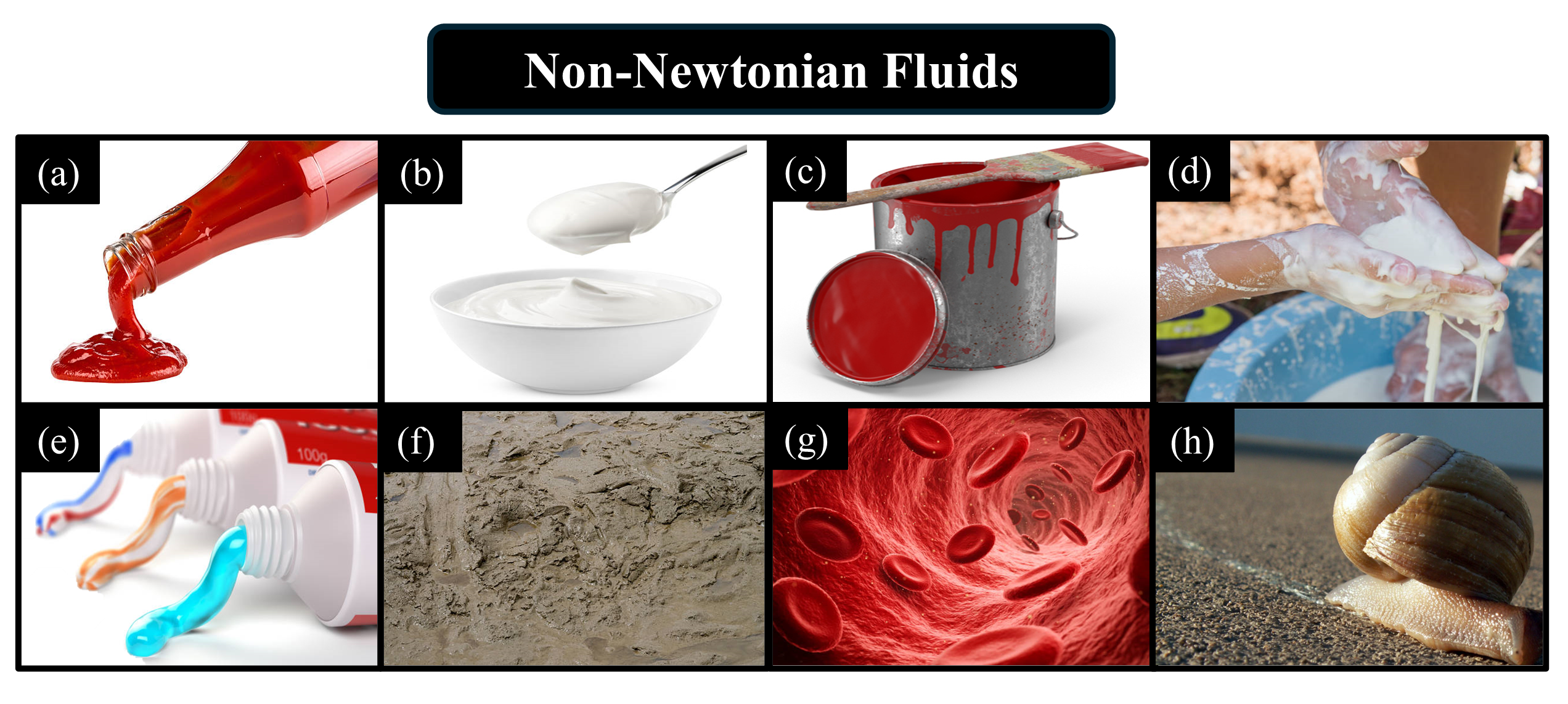}
	\caption{Illustration of common non-Newtonian fluid examples, which encompass various types of anomalous flow behaviour. These involve shear-thinning in (a) ketchup and (b) yoghurt, as well as thixotropy in (c) paints, while shear-thickening is observed in (d) a suspension of cornstarch and water (commonly referred to as Oobleck). Viscoplasticity is observed in materials that exhibit a yield-stress transition, such as (e) squeezing toothpaste and (f) mud. An additional elastic material response can be naturally observed in certain biofluids, such as (g) blood and (h) mucus. Note, certain examples outlined above, such as ketchup, can indeed experience multiple anomalous flow behaviours, such as shear-thinning and viscoplasticity (i.e., shaking the ketchup bottle causes it to flow from rest and reduce its viscosity).   }
	\label{figure_1}
\end{figure*}

Non-Newtonian fluids are ubiquitous in nature and everyday life, occurring in a variety of
cosmetics, foods, natural substances, and biological fluids. Such fluids can be further categorised based on their anomalous flow behaviour. For instance, several examples highlighted in Fig.~\ref{figure_1}, such as ketchup and yoghurt, exhibit shear-thinning qualities, whereby the viscosity decreases as the shear rate increases, facilitating easier flow under stress. Additionally, these fluids can often exhibit thixotropy --- a time-dependent behaviour in which the viscosity gradually decreases over time under a constant shear rate \cite{MEWIS19791,BARNES19971,Larson_Thixo}. In contrast, shear-thickening fluids, like cornstarch suspensions (i.e., Oobleck), become more viscous when subjected to higher shear rates, often leading to counterintuitive and abrupt flow resistance, as further described in several recent reviews \cite{Morris_Shear_Thick,Tian_Shear_Thick, WEI2022110570,GURGEN201748}. In addition to nonlinear viscous effects, certain fluids can exhibit additional material properties, such as plastic or solid-like behaviour, commonly referred to as viscoplastic or yield-stress fluids \cite{Yield_Stress_Rev,COUSSOT201431,YIELD_STRESS_RHEOL_ACTA}. Intuitive examples of such materials include toothpaste and mud, which are governed by a threshold stress (i.e., yield-stress criteria), above which they flow as fluids and below they behave as solids. Furthermore, many complex fluids, such as polymer solutions, suspensions, emulsions, and biofluids (i.e., blood, mucus, saliva, etc.), exhibit viscoelasticity, generating an additional linear or nonlinear elastic contribution to the total fluid stress \cite{Viscoelastic_ISSUES_REV,Datta_Pers}.
\begin{figure*}[t]
	\centering
	\includegraphics[width=0.8\textwidth]{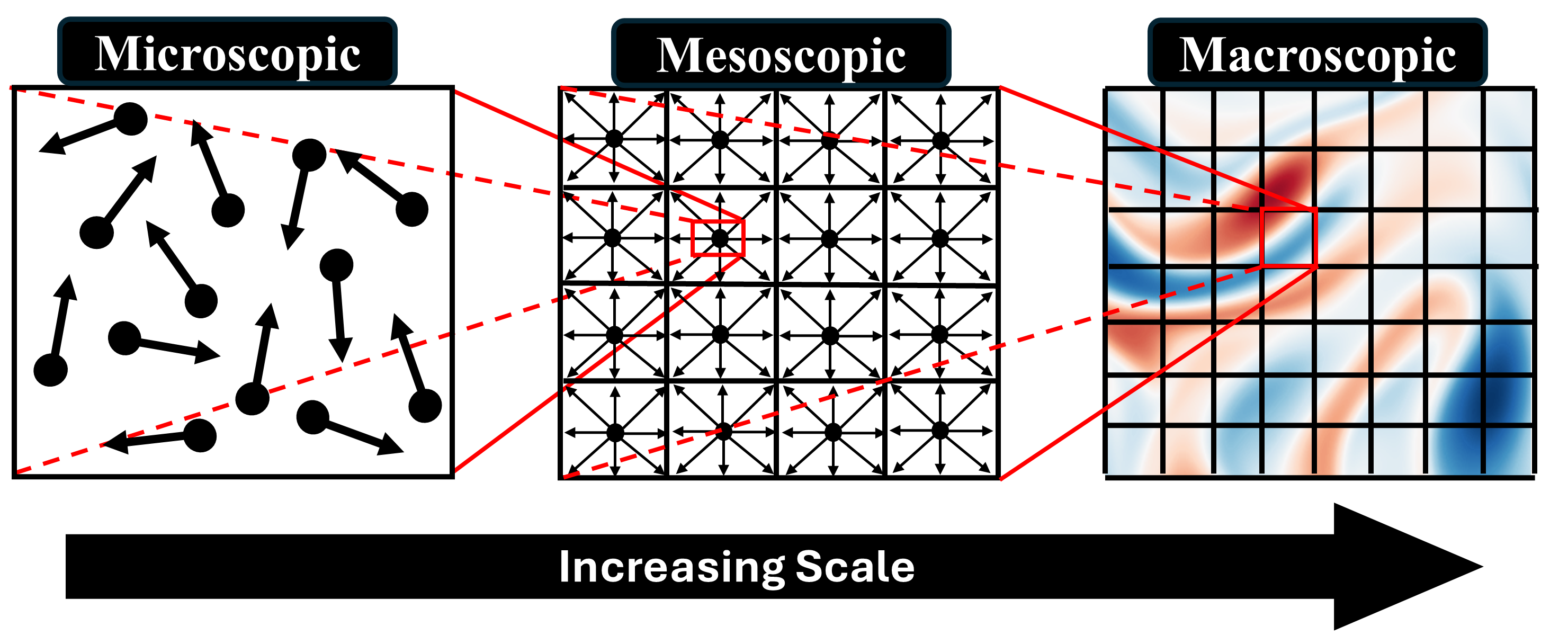}
	\caption{Example of computation fluid dynamics (CFD) techniques applied across different scales: (from left to right) microscopic, mesoscopic, and macroscopic. Each resolution highlights the distinct approaches used to capture fluid behaviour, ranging from detailed molecular interactions at the microscopic scale [i.e., molecular dynamics (MD)], to particle-based models at the mesoscopic scale [i.e., lattice Boltzmann method (LBM), smoothed-particle hydrodynamics (SPH), dissipative particle dynamics (DPD), etc.], and continuum-based equations at the macroscopic scale (i.e., NSE). The transitions emphasize the trade-offs in computational complexity and the level of physical detail represented at each scale. Although microscopic simulations provide unparalleled detail, they are computationally expensive, while macroscopic models prioritise efficiency at the expense of fine-grained details.}
	\label{figure_2}
\end{figure*}

Understanding and describing the behaviour of different non-Newtonian systems remains a challenging task. The inherent difficulties in developing
generalised mathematical formulations for the behaviour of any non-Newtonian fluid system becomes increasingly complicated when considering the aforementioned enormous range of different anomalous flow behaviours occurring in non-Newtonian fluids (refer to Fig.~\ref{figure_1}). A common approach to modelling non-Newtonian fluid behaviour involves varying the viscosity term based on the shear rate at a given moment, referred to as the effective viscosity. These generalized Newtonian fluids (GNFs) assume that the viscosity depends only on the instantaneous shear rate and is independent of the deformation history \cite{Irgens2014,PEARSON2002447,POOLE2023105106,OWOLABI2023105089}. Popular approaches include the power-law and Carreau–Yasuda models, which are able to describe shear-thinning or shear-thickening behaviour based on tuning empirically derived parameters \cite{Irgens2014}. Certain GNF's, such as the Bingham and Herschel–Bulkley models, impose an additional yield stress criterion to describe viscoplastic behaviour \cite{Yield_Stress_Rev,ZHU2005177}. However, when deformation history plays an important role, particularly in viscoelastic systems, other non-Newtonian models beyond GNFs are required \cite{Irgens2014,POOLE2023105106,OWOLABI2023105089}. At relatively small deformations (i.e. small strains and small rotations), one can resort to simple linear viscoelastic models, such as the Maxwell model, which is capable of predicting stress relaxation and creep. However, at higher deformations, more complex nonlinear constitutive models, such as the Oldroyd-B \cite{Oldroyd1950OnTF} or FENE-P models \cite{PETERLIN1961257}, are required to predict strong elastic features, including normal stress differences \cite{Irgens2014, Bird_Book}.

Combined, the above GNFs and constitutive models can describe a variety of anomalous non-Newtonian flow behaviours. In simple, but practically important flow problems, such as Hagen-Poiseuille flow \cite{SCHLEINIGER199179,oliveira_pinho_1999,Cruz_2005,Rahmani_Taghavi_2022}, plane Couette flow \cite{DAPRA2015221,Couette_Sol,ALIBENYAHIA201237}, and creeping flow past cylinders \cite{Bruschke,WOODS2003211}, to name a few, these models can indeed be solved analytically. However, given the propensity for non-Newtonian fluids to occur in practically complex problems, as well as the requirement to solve additional non-trivial constitutive equations when dealing with viscoelasticity, such problems necessitate the use of numerical methods based on computational fluid dynamics (CFD) \cite{Irgens2014,Bird_Book}. Typically, the Navier-Stokes equations (NSE) coupled with the appropriate non-Newtonian model are solved directly based on a continuum-based assumption (Fig.~\ref{figure_2}), using conventional CFD solvers, such as finite difference schemes, finite volume methods, or finite element approaches \cite{Alves_Rev,DRITSELIS2022105590,MarnDelicZunic+2001+325+335,Punia,Muhammad_NN,HASSAGER1983153,huang2012finite}. Due to their relative simplicity and long-standing tradition, these conventional solvers are typically viewed as workhouse methods in CFD, being widely implemented in a variety of open-source production flow solvers \cite{OpenFoam,PIMENTA201785}, generally with second-order accuracy \cite{Krueger}. However, their reliance on continuum assumptions limits their applicability to problems where molecular-level interactions or complex multiscale phenomena are critical, such as in high Knudsen number flows \cite{LBM_MICRO}. 

Microscopic methods, such as molecular dynamics (MD), overcome this limitation by simulating individual particle interactions (refer to Fig.~\ref{figure_2}), enabling high-fidelity simulations of non-Newtonian behaviour arising from the fluid's microstructural dynamics \cite{SCI_REP_MOL,CHEN2017378,BUSIC200351}. Despite offering unmatched detail, the computational expense of these methods scales poorly with system size, making them impractical for large-scale or time-intensive simulations even with the most advanced supercomputers \cite{Kadau,hollingsworth2018molecular}. Mesoscopic approaches emerge as an ideal compromise, providing an intermediate avenue towards capturing the full scale of the problem (Fig.~\ref{figure_2}). These particle-based approaches are fundamentally grounded in kinetic theory principles, focusing on modelling the collective behaviour of particle groups as they evolve over time \cite{Krueger,LBM_MICRO}. Popular approaches include dissipative particle dynamics (DPD) \cite{DPD_PRE,DPD_Chem,DPD_ARCH}, smoothed-particle hydrodynamics (SPH) \cite{SPH,Benz1990,SPH_Lind}, and the lattice Boltzmann method (LBM) \cite{Krueger,LBM_MICRO,book_LBM,Succi2001,shan_2006,Aidun}. 

DPD is a mesh-free, coarse-grained extension of MD, where each particle represents a collection of molecules. Interactions between particles are governed by conservative, dissipative, and random forces \cite{DPD_ARCH}. However, DPD relies on a large number of parameters whose values do not directly correspond to macroscopic quantities of interest and have to be selected carefully to avoid unphysical flow behaviour \cite{Krueger,Filip}.  SPH is an alternative mesh-free, Lagrangian particle-based method that models fluids by tracking the movement of individual particles and their interactions. At the heart of SPH is an interpolation scheme which uses point particles that
influence their vicinity \cite{SPH}. 
While SPH has gained significant popularity in the astrophysics community over the past few decades \cite{SPH,SPH_Lind,Springel}, it is known to face challenges such as unwanted compressibility effects, high computational costs, and difficulties in managing boundary conditions \cite{Krueger}. 

LBM represents a distinct approach within particle-based methods, bridging the microscopic and macroscopic scales through a discretized Boltzmann equation. Unlike DPD and SPH, LBM operates on a fixed, typically structured, grid, evolving particle distribution functions rather than individual particles \cite{Krueger,Succi2001,shan_2006}. This framework, which includes highly localised operations, is well-suited for parallel processing, making it ideal for implementation on GPU-accelerated architectures \cite{GPU_LBM,GPU_O,GPU_RIN,GPU_XIAN}. That being said, LBM, like any other CFD solver, is not without its own flaws and limitations \cite{Krueger,book_LBM}, as will be discussed later on. Notwithstanding, certain key features, such as the simplicity in implementing boundary conditions \cite{LBM_BC_1,LBM_BC_2,LBM_BC_3}, modelling multiphase interactions without explicitly interface tracking \cite{Shan1995MulticomponentLM,Martys_PRE,Wang2019,CHEN2014210,Huang_2011,Li_2012}, and ease of including additional multiphysics \cite{Aidun,TIRIBOCCHI20251,Carenza2019,huang2021transition,huang2024self} make LBM particularly attractive for applications involving complex flow problems. Thus, it has become increasingly popular to extend LBM to simulate different anomalous non-Newtonian flow behaviours \cite{Phillips_2011, LBM_BLOOD,LBM_BLOOD_COMP,LBM_NN_POROUS,Gupta_Hybrid,DZANIC2022105280}. Notably, the review by Phillips \& Roberts \cite{Phillips_2011} provided a thorough overview of the different techniques applied to LBM to simulate two major classes of non-Newtonian fluids: inelastic fluids characterised by a variable viscosity and viscoelastic fluids characterised by fluid elasticity. Over the past decade, several refinements have been made to these extended LBMs, which have enabled their application to more realistic and challenging problems involving non-Newtonian fluids. Here, we provide a comprehensive review of these LBM advancements for describing various anomolous non-Newtonian behaviours, specifically, shear-thinning, shear-thicking viscoplasticity, and viscoelasticity.

The remainder of this review paper is organised as follows. In Section~\ref{sec:LBM}, we first formally introduce the general LBM and its fundamental algorithm process, which is then extended to include additional multiphysics. To accommodate for readers who are not well-versed with LBM, we conclude the section by briefly discussing the main benefits and limitations of LBM. In Section~\ref{sec:GNF}, we review different LBM techniques and applications for simulating non-Newtonian flow behaviours described by GNFs, specifically shear-thinning, shear-thickening, and viscoplasticity, concluding with a discussion of the key challenges and future directions in this area. In Section~\ref{sec:Viscoelastic}, we provide a detailed review of previous attempts to incorporate viscoelasticity into LBM, along with their recent applications to challenging problems, again highlighting remaining challenges and perspectives for future research. Finally, Section~\ref{sec:Conclusions} presents overall conclusions.

\section{The Lattice Boltzmann Method}\label{sec:LBM}
\subsection{LBM for Simple Hydrodynamics}\label{SS:LBM_Basics}

\begin{figure}
	\centering
	\includegraphics[scale=0.6]{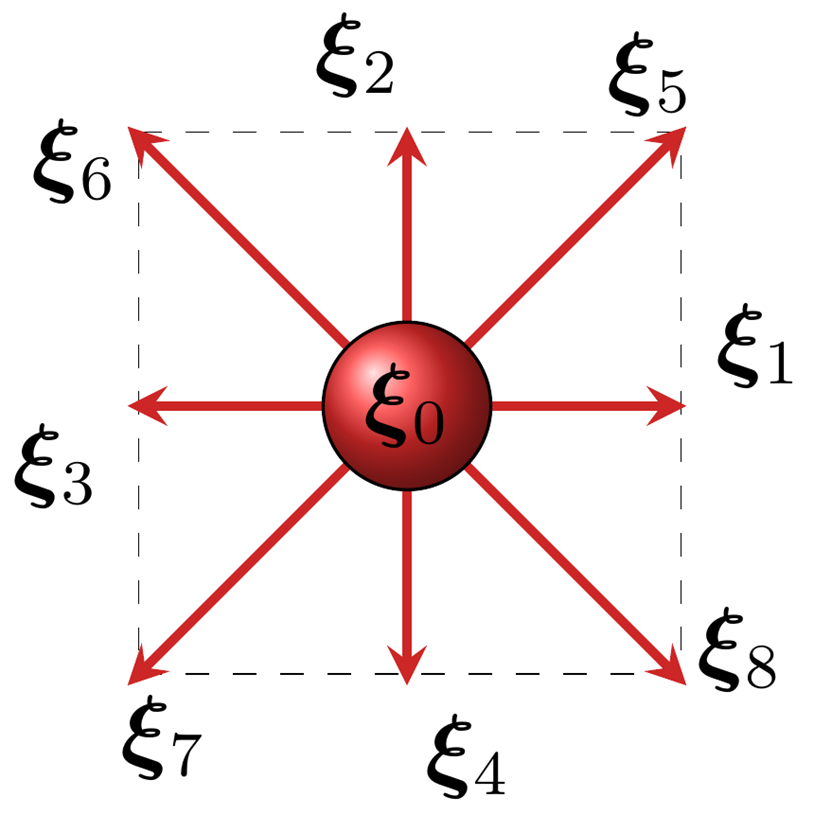}
	\caption{The two-dimensional nine-velocity lattice structure set $\{ w_{\alpha}, \bm{\xi}_{\alpha} : \alpha=0,\dots, 8 \}$, known as the D2Q9, where $c_s^2=1/3$.}
	\label{figure_3}
\end{figure}

\noindent An alternative to describing the hydrodynamic field with the continuum-based NSE is the kinetic theory approach offered by LBM. In this framework, the fluid behaviour is instead described by a phase-space discretised form of the Boltzmann equation. In doing so, fluids are instead described by fictitious particle groups using distribution functions, $f_{\alpha}$, which are discretised by projecting a predefined lattice structure set comprising of discrete distribution weights ($ w_{\alpha} $) and lattice velocities ($ \bm{\xi}_{\alpha} $), where $ c_s^2 \bm{I} = \sum_{\alpha}w_{\alpha}\bm{\xi}_{\alpha}\bm{\xi}_{\alpha}$ is the lattice-structure dependent sound speed. Akin to the variety of discretisation schemes offered in conventional CFD, LBM features different lattice structures $D_{d}Q_{q}$ dependent on the spatial dimension $d$ and the number discrete terms $q$, which are tailored to specific applications \cite{Qian_1992}. For instance, popular examples in the literature generally used to resolve the NSE hydrodynamic limit include the D2Q9 (see Fig.~\ref{figure_3}), and its extension to 3D with D3Q15, D3Q19, and D3Q27. Notably, it is indeed possible to construct higher-order lattice structures, capable of describing hydrodynamic details beyond the NSE capabilities \cite{shan_2006,Chikatamarla,dzanic2020investigation}, playing particular importance in simulating non-equilibrium phenomena at high Knudsen numbers \cite{LBM_MICRO,PRE_HIGHORDER}. However, for simplicity, we will only focus on the D2Q9 lattice structure here, which can be defined by the discrete velocity set,  
\begin{equation}
\bm{\xi}_{\alpha} =
\begin{cases} 
(0, 0), & \alpha = 0, \\ 
(1, 0), (-1, 0), (0, 1), (0, -1), & \alpha = 1, 2, 3, 4, \\ 
(1, 1), (-1, 1), (-1, -1), (1, -1), & \alpha = 5, 6, 7, 8,
\end{cases}
\end{equation}
and corresponding lattice weights,
\begin{equation}
w_{\alpha} =
\begin{cases} 
\frac{4}{9}, & \alpha = 0, \\ 
\frac{1}{9}, & \alpha = 1, 2, 3, 4, \\ 
\frac{1}{36}, & \alpha = 5, 6, 7, 8,
\end{cases}
\end{equation}
which ensure the required level of isotropy and accuracy in recovering the NSE \cite{Krueger}. 

The portion of particles moving with $\alpha$th lattice velocity at lattice site $x$ and time $t$ [i.e., $f_{\alpha}(x,t)$] can be modelled using the discretised Boltzmann transport equation,
\begin{multline}\label{eq:BGK}
	f_{\alpha} \left(\bm{x}+\bm{\xi}_{\alpha} \Delta t,t+\Delta t \right) = f_{\alpha}(\bm{x},t) + \Omega_{\alpha}^{\rm{SRT}}(\bm{x},t),
\end{multline}
where $f_{i} \left(\bm{x}+\bm{\xi}_{\alpha}\Delta t,~ t+\Delta t \right)$ is known as the streaming step (i.e., advection), which involves propagating particle distributions to their nearest lattice neighbours, i.e., ($\bm{x}+\bm{\xi}_{\alpha} \Delta t$). Whereas, the collision operator, 
\begin{equation}\label{eq:BGK_Collision}
	\Omega_{\alpha}^{\rm{SRT}}(\bm{x},t) = -\frac{\Delta t}{\tau} \left[f_{\alpha}(\bm{x},t) - f_{\alpha}^{eq}(\bm{x},t) \right],
\end{equation}
simply relaxes the particle populations towards a local equilibrium $f_{\alpha}^{eq}(\bm{x},t)$, at a rate determined by the Bhathagar-
Gross-Krook (BGK) single relaxation time (SRT) approximation, $\tau$ \cite{BGK_1954}. Importantly, $\tau$ is physically related to the kinematic shear viscosity $\nu=c_s^2(\tau-\frac{1}{2})\frac{\Delta x^2}{\Delta t}$. Here, $\Delta x$ and $\Delta t$ represent the lattice spacing and time step, respectively, and are typically set to unity for convenience, i.e., $\Delta x = \Delta t = 1$.
The equilibrium term $f_{\alpha}^{eq}(\bm{x},t)$ represents the discretised Maxwell-Boltzmann equilibrium distribution truncated up to second-order Hermite polynomials \cite{shan_2006},
\begin{equation}\label{eq:Feq}
	f_{\alpha} ^{eq} = w_{\alpha} \rho \biggl\{ 1 + \underbrace{ \frac{\bm{\xi}_{\alpha} \bm{\cdot} \textcolor{black}{\bm{u}^{eq}}}{c_s^2}  }_\text{$1$st Order} + \underbrace{\frac{1}{2}\left[\left(\frac{\bm{\xi}_{\alpha} \bm{\cdot} \textcolor{black}{\bm{u}^{eq}}}{c_s^2}\right)^2 - \frac{(\textcolor{black}{{u}^{eq}})^2}{c_s^2} \right]}_\text{$2$nd Order}\biggr\},
\end{equation}
\textcolor{black}{where $\bm{u}^{eq}$ is the equilibrium velocity.} Note, for typographical convenience, explicit spatial and temporal dependence have been omitted above.

Despite operating in the mesoscopic framework, key macroscopic variables, such as density $\rho$ and velocity $\bm{u}$, are readily obtained through simple moment summation operations of $f_{\alpha}$,
\begin{equation}
    \rho = \sum_{\alpha} f_{\alpha},
\end{equation}
\begin{equation}
    \rho \bm{u} = \sum_{\alpha} f_{\alpha} \bm{\xi}_{\alpha}. 
\end{equation}
Based on the LBM scheme described, a direct link back to the macroscopic NSE can be obtained through Chapman-Enskog expansion (for full derivation see \cite{Krueger,chapman1990mathematical}), retrieving,
\begin{equation}\label{eq:NSE}
    \bnabla \bcdot \bm{u} = 0, ~~~ \frac{\partial \bm{u}}{\partial t} + \bm{u} \cdot \bnabla \bm{u} = - \frac{1}{\rho} \nabla p + \nu \bnabla^2 \bm{u}.
\end{equation}
To briefly summarise the standard LBM scheme, Eq.~(\ref{eq:BGK}) can be presented in a more convenient manner, which directly coincides
with the key steps in the LBM algorithm process. Whereby, the algorithm consists of the following
three fundamental steps to describe the evolution of particle groups,
\begin{enumerate}
    \item \textbf{Collision Step:} 
    \[
    f^*_{\alpha}(\bm{x},t) = f_{\alpha}(\bm{x},t) + \Omega_{\alpha}^{\rm{SRT}}(\bm{x},t).
    \]
    \item \textbf{Streaming Step:}
    \[
    f_{\alpha} \left(\bm{x}+\bm{\xi}_{\alpha} \Delta t,t+\Delta t \right) = f^*_{\alpha}(\bm{x},t).
    \]
    \item \textbf{Obtain Hydrodynamic Moments:}
    \[
    \rho(\bm{x}, t) = \sum_{\alpha} f_{\alpha}(\bm{x}, t).
    \]
    \[
    \bm{u}(\bm{x}, t) = \frac{1}{\rho(\bm{x}, t)} \sum_{\alpha}f_{\alpha}(\bm{x}, t) \bm{e}_{\alpha}.
    \]
\end{enumerate}
In practical implementations, macroscopic boundary conditions, most notably no-slip walls, are mesoscopically enforced post-streaming using bounce-back schemes, which provide second-order accuracy on uniform grids and remain widely used due to their simplicity and robustness \cite{Krueger, Succi2001}.

It is important to note that there are several other more complicated LBM schemes proposed for fluid dynamics \cite{Krueger}, which will not be covered here. These include more advanced collision operators, such as multiple relaxation time (MRT) schemes \cite{Humieres,Lallemand_MRT}, higher-order LBMs \cite{shan_2006,Chikatamarla}, and unstructured LBM models \cite{Ubertini,Rossi_LBM}. Whereby, MRT schemes are extensions to the BGK SRT in Eq.~(\ref{eq:BGK_Collision}), which instead perform the collision step in momentum space, providing, in certain cases, enhanced numerical stability and accuracy, while also offering the option to adjust the bulk and
shear viscosities independently \cite{McCracken,YOSHIDA20107774,Ginzburg}. However, the careful tuning of multiple relaxation rates is far from trivial, as it is often highly case-dependent and requires meticulous attention to the specific flow conditions and problem parameters \cite{Krueger,Chiappini}. The full technical details of these models is beyond the current scope and can be found in previous works \cite{Krueger,Humieres,Lallemand_MRT}.

\subsection{Additional Momentum Contributions in LBM}\label{SS:Forces}
\noindent Forces play a central role in many hydrodynamic problems. The inclusion of additional internal or external momentum contributions into LBM, resulting from gravity, non-ideal interactions, electric or magnetic forces, etc., is relatively straightforward. This is typically achieved by adding a forcing term $\Omega_{\alpha}^{F}(\bm{x},t)$ during the LBM collision step \cite{He_1997,Guo_2002},
\begin{equation}\label{eq:BGK_Force}
	f_{\alpha} \left(\bm{x}+\bm{\xi}_{\alpha} \Delta t,t+\Delta t \right) = f_{\alpha}(\bm{x},t) + \Omega_{\alpha}^{\rm{SRT}}(\bm{x},t)
 + \Omega_{\alpha}^{F}(\bm{x},t).
\end{equation}
An external forcing contribution $\bm{F}$ can be coupled explicitly with $ f_{\alpha}^{eq} $ to conserve the explicit nature of LBM using different second-order accurate forcing schemes \cite{LB_FORCE_REV}. These different variations for $\Omega_{\alpha}^{F}(\bm{x},t)$ have been previously compared in literature \cite{Huang_2011,Li_2012,DZANIC_FEEDBACK}. Here, we describe two popular approaches, namely, the explicit forcing (EF) scheme by He \textit{et al.}  \cite{He_1997} and the Guo \textit{et al.} scheme \cite{Guo_2002}. Whereby, the EF scheme assumes that equilibrium contributions play a major role in the continuous LBM forcing term, thus obtaining a discrete approximation,
\begin{equation}\label{eq:EF_Scheme}
	\Omega_{\alpha}^{F}(\bm{x},t)=\Delta t \left(1 - \frac{1}{2\tau} \right)\frac{ \bm{F}\bm{\cdot} \left(\bm{\xi}_{\alpha}-{\bm{u}^{eq}} \right)}{\rho c_s^2}f_{\alpha}^{eq},
\end{equation}
where the equilibrium velocity, $\bm{u}^{eq} = \sum_{\alpha} f_{\alpha} \bm{\xi}_{\alpha} + \frac{\Delta t}{2\rho} \bm{F}$, is used to compute $f_{\alpha}^{eq} $ in Eq.~(\ref{eq:Feq}). 

Similarly, the forcing scheme proposed by Guo \textit{et al.} \cite{Guo_2002}, strategically derived through Chapman-Enskog expansion to retrieve the exact NSE representation up to second-order, has the form,
\begin{equation}\label{eq:Guo_Scheme}\resizebox{.99\hsize}{!}{$
	\Omega_{\alpha}^{F}(\bm{x},t)=\Delta t \left(1 - \frac{1}{2\tau} \right)w_{\alpha}\left(\frac{1}{c_s^2}\left(\bm{\xi}_{\alpha}-\textcolor{black}{\bm{u}^{eq}} \right)+\frac{1}{c_s^4} \xi_{\alpha}^2\textcolor{black}{\bm{u}^{eq}}\right)\bm{\cdot}\bm{F}$},
\end{equation}
where $\xi_{\alpha}^2 = \bm{\xi}_{\alpha}\bm{\cdot}\bm{\xi}_{\alpha}$, the equilibrium velocity, $\bm{u}_{eq} = \sum_{\alpha} f_{\alpha} \bm{\xi}_{\alpha} + \frac{\Delta t}{2\rho} \bm{F}$, and used to compute $f_{\alpha}^{eq} $ in Eq.~(\ref{eq:Feq}).

It can be shown that the modifications in these forcing schemes increase the fluid momentum at a lattice node by an amount of $\frac{\Delta t}{2}\bm{F}$ during
each $\Delta t$ without changing the fluid density,
\begin{equation}\label{F1}
	\rho = \sum_{\alpha} f_{\alpha}, 
\end{equation}
\begin{equation}\label{F2}
	\rho \bm{u} = \sum_{\alpha} f_{\alpha} \bm{\xi}_{\alpha} + \frac{\Delta t}{2} \bm{F}.
\end{equation}  

An alternative approach also exists to instead include additional momentum contributions directly through the second-order moment of $f_{\alpha}^{eq}$ in Eq.~(\ref{eq:Feq}) \cite{Swift_1996,Holdych1998AnIH}, such that,
\begin{multline}\label{eq:Feq_Mod}
	f_{\alpha} ^{eq} = w_{\alpha} \rho \biggl\{ 1 + \frac{\bm{\xi}_{\alpha} \bm{\cdot} \textcolor{black}{\bm{u}^{eq}}}{c_s^2} + \frac{1}{2}\left[\left(\frac{\bm{\xi}_{\alpha} \bm{\cdot} \textcolor{black}{\bm{u}^{eq}}}{c_s^2}\right)^2 - \frac{(\textcolor{black}{{u}^{eq}})^2}{c_s^2} \right]\biggr\} \\
    -w_{\alpha}\left(\frac{\bm{\xi}_{\alpha} \bm{\xi}_{\alpha}}{2c_s^4}-\frac{\bm{I}}{2c_s^2}\right)\bm{:}\bm{\Pi},
\end{multline}
where ``$\bm{:}$'' denotes full tensor contraction. This approach, in turn, includes additional contributions to the symmetric part of the total stress tensor, and hence, by design only works for external contributions $\bm{F}=\bnabla\bcdot\bm{\Pi}$, where the rank-2 stress tensor $\bm{\Pi}$ is symmetric \cite{Denniston}. In doing so, the following hydrodynamic variables are obtained as moments of $f_{\alpha}$,
\begin{equation} \label{F4}
	\rho = \sum_{\alpha} f_{\alpha},
\end{equation}
\begin{equation} \label{F5}
    \rho \bm{u} = \sum_{\alpha} f_{\alpha} \bm{\xi}_{\alpha},
\end{equation}
\begin{equation} \label{F6}
    \rho c_s^2 \bm{I} + \rho\bm{u}\bm{u} + \bm{\Pi}= \sum_{\alpha} f_{\alpha} \bm{\xi}_{\alpha} \bm{\xi}_{\alpha},
\end{equation}
where it can be seen that the additional stress $\bm{\Pi}$ is directly implemented within the second-order moment as an additional contribution to the total momentum flux, which includes the equilibrium pressure $P=\rho c_s^2$ and kinetic term $\rho\bm{u}\bm{u}$.
  
\subsection{Multiphase and Multicomponent flows with LBM} \label{SS:Multiphase}
%
%
\noindent In certain cases, fluid systems may exhibit multiple phases or consist of miscible or immiscible mixtures composed of multiple components. In this subsection, we briefly discuss popular approaches for extending LBM to include these additional non-ideal interactions.

The flexibility of the source term in LBM allows for various complex multiphase (see, e.g., \cite{Li_KH-Luo_etal_2016_REVIEW} for a comprehensive review) and multicomponent flows to be simulated efficiently \cite{Succi_2015}. 
For instance, single-component multiphase flows include a non-ideal force term described by a thermodynamic equation of state.
Multiple components ($\phi=A,~B$) can be simulated, each with its own discrete distribution function $f_{\alpha}^{\phi}:\phi=A,B$ [Eq.~(\ref{eq:BGK_Force})], i.e.,
\begin{equation}\label{eq:EF__Scheme_Multi}
f_{\alpha}^{\phi} \left(\bm{x}+\bm{\xi}_{\alpha},t+\Delta t \right) = f_{\alpha}^{\phi}(\bm{x},t) - \Omega_{\alpha}^{\mathrm{SRT},\phi}(\bm{x},t)  +  \Omega_{\alpha}^{F,\phi}(\bm{x},t),
\end{equation}
where the components are then coupled via the source term ($ \Omega_{\alpha}^{F,\phi} $) and the collision operator ($ \Omega_{\alpha}^{\mathrm{SRT},\phi} $) to ensure the conservation of mass and momentum of the entire system. For each $\phi$-component, the collision $ \Omega_{\alpha}^{\mathrm{SRT},\phi} $ relaxes towards $\tau^{\phi}$ based on their individual kinematic shear viscosity $\nu^{\phi}=c_s^2(\tau^{\phi}-\frac{1}{2})\frac{\Delta x^2}{\Delta t}$.
The hydrodynamic variables of each $\phi$-component are defined from the moments of $f_{\alpha}^{\phi}$, e.g., the density (\ref{F1}) and momentum (\ref{F2}),
\begin{equation*}
\rho^{\phi}=\sum_{\alpha}^{} f_{\alpha} ^{\phi}, ~~ \text{and}~~
\rho^{\phi}\bm{u}^{\phi} =\sum_{\alpha}^{} f_{\alpha} ^{\phi} \bm{\xi}_{\alpha} + \frac{\Delta t}{2}\bm{F}^{\phi}.
\end{equation*}
%
%
Non-ideal forces ($\bm{F}^{\phi}$) are then introduced via an extension.

There exist various approaches to include non-ideal forces \cite{Krueger,Succi_2018}, notably the \textit{free-energy model} \cite{Swift_1996} and the \textit{pseudopotential model} \cite{Shan_Chen_1993}. Other methods, such as the \textit{colour-gradient model} \cite{Gunstensen_1991_PhysRevA.43.4320}, introduce a surface tension by an additional nonlinear collision perturbation term after the collision step, rather than a force term, followed by a recolour step that enforces a sharp interface; however, fluids are inherently segregated (i.e., cannot account for mixed regions).

In the free energy model by Swift~\textit{et~al.~}\cite{Swift_1996}, a thermodynamic pressure tensor is derived from a functional of the Helmholtz free energy (known as the free-energy functional), described by a non-ideal equation of state, such as the van der Waals equation of state. 
The free energy model can be incorporated into the collision [Eq.~(\ref{eq:BGK_Force}) or Eq.~(\ref{eq:EF__Scheme_Multi})] as non-ideal forces through $F_{i}=-\nabla\cdot (P_{ij} - c_s^2\rho I_{ij})$ \cite{Pooley_Furtado_2008_PhysRevE,Wen_PRE_2017,Wen_PRE_2020,Soomro_PRE_2023}.
More specifically, the gradient of chemical potential $\mu$, derived from the free-energy functional, can be shown to be related to the divergence of the pressure tensor as $\nabla\cdot P_{ij} = \rho\nabla\mu$ (see \cite{Wen_PRE_2017}), such that the non-ideal force,
\begin{equation}\label{eq:FreeEnergy_}
\bm{F}(\bm{x})=-\rho(\bm{x})\nabla \mu(\bm{x}) + c_s^2 \nabla \rho(\bm{x}).
\end{equation}
The above non-ideal force is relevant only at interfaces (in the presence of density gradients), such as in single-component multiphase or immiscible multicomponent flows. Miscible multicomponent systems traditionally require a bulk chemical potential defined from a multicomponent free-energy functional, which is not readily available. Recent work by Soomro~\textit{et~al.}~\cite{Soomro_PRE_2023} addressed this by deriving a bulk chemical potential formulation based on fugacity property. Moreover, similar external forces, such as a chemical potential gradient normal to surfaces (boundaries), which, in turn, supplement the bounce-back boundary conditions, can be used to alter surface wettability \cite{Swift_1996}.

The Shan-Chen pseudopotential interaction model \cite{Shan_Chen_1993} provides a unique picture of interactions in fluids by mimicking intermolecular force interactions \textit{discretely} on the lattice set $\{ \tilde{w}_{\alpha}, \bm{\xi}_{\alpha} \}$, specifically self-interactions (${G}^{\phi\phi}, \psi^{\phi}:\phi = A,B$) and cross-interactions (${G}^{AB}$, $\Psi^{A}$, and $\Psi^{B}$), i.e., 
\begin{equation}\label{eq:SC_self_cross_interactions}
\begin{aligned}
	\bm{F}^{\phi}(\bm{x}) = 
	- & \psi^{\phi}(\bm{x}) {G}^{\phi\phi} \sum_{\alpha}^{} \tilde{w}_{\alpha}\psi^{\phi}(\bm{x}+\bm{\xi}_{\alpha})\bm{\xi}_{\alpha} 
	\\
	- & \Psi^{\phi}(\bm{x}) {G}^{\phi\varphi} \sum_{\alpha}^{} \tilde{w}_{\alpha}\Psi^{\varphi}(\bm{x}+\bm{\xi}_{\alpha})\bm{\xi}_{\alpha},
\end{aligned}
\end{equation}
where superscript `$\varphi$' denote any other component $\varphi\neq\phi$ and symmetry require ${G}^{\phi\varphi}={G}^{\varphi\phi}$. The spatial accuracy of (\ref{eq:SC_self_cross_interactions}) conforms to the isotropic gradients (rotational invariance) of the lattice set $\{\tilde{w}_{\alpha},\bm{\xi}_{\alpha}\}$ due to its discrete construction on the lattice \cite{Chris_2019,Chris_2020,Lulli_PRE_2021}.
To model immiscible fluids, the pseudopotential interaction force takes the form,
\begin{equation}\label{eq:SC_multi}
\bm{F}_{int}^{\phi}\left(\bm{x}\right)=
-\psi^{\phi}\left(\bm{x}\right)\sum_{\widetilde{\phi}\neq\phi}^{S}G^{\phi\widetilde{\phi}}\sum_{\alpha}w_{\alpha}\psi^{\widetilde{\phi}}\left(\bm{x}+\bm{\xi}_{\alpha}\Delta t\right)\bm{\xi}_{\alpha}\Delta t.
\end{equation}
where $\psi^{\phi}$ is the “effective” density function for the $\phi$ component, and $G^{\phi\widetilde{\phi}}>0$ controls the repulsive interaction strength between immiscible components. 
Similarly, wettability can be readily implemented by considering an additional interaction force between different fluid components $\phi$ with the surrounding solid boundaries, $\bm{x}^{s}$,
\begin{equation}\label{eq:SC_surface}
\bm{F}^{s,\phi}=-G^{s\phi}\psi^{\phi}\left(\bm{x}\right)\sum_{\alpha}w_{\alpha}\psi^{s,\phi}\left(\bm{x}+\bm{\xi}_{\alpha}\Delta t\right)\bm{\xi}_{\alpha}\Delta t,
\end{equation}
where $\psi^{s,\phi}\left(\bm{x}+\bm{\xi}_{\alpha}\Delta t\right)$ is a binary indicator equal to 1 or 0 for a solid or a fluid node, respectively. The interaction strength $G^{s\phi}$ between fluid components $\phi$ with $\bm{x}^{s}$ can be adjusted accordingly, where positive (negative) values reflect nonwetting (wetting) conditions \cite{Huang_PRE,Chris_2020}. 

\subsection{Summary---Why LBM?}\label{sec:Why_LBM}
\noindent Having outlined the technical details of LBM, it is worth highlighting why this approach has gained growing attention for solving diverse flow problems, including non-Newtonian systems. Put simply, the growing shift towards LBM emerging as a conventional CFD solver stems from its unique characteristics. For instance, LBM derives a major advantage from being based on the Boltzmann equation rather than the NSE, making it significantly simpler to implement compared to conventional methods. As described in Subsection~\ref{SS:LBM_Basics}, LBM follows a relatively straightforward algorithm process consisting of simple streaming and collision steps to evolve the flow field, whereby key hydrodynamic quantities can be readily obtained through simple summation operations. In addition to LBM's simplicity, the collision step itself is fully localised, making LBM very amenable to high-performance computing on parallel architectures, including GPUs \cite{GPU_LBM,GPU_O, GPU_RIN,GPU_XIAN}. With continual advancements in computational technology, particularly in GPU capabilities and parallel processing power \cite{schornbaum2016massively,schmieschek2017lb3d,latt2021palabos}, LBM is positioned as a pivotal tool for tackling increasingly complex and large-scale fluid dynamics problems. 

The streaming step itself also offers unique advantages: the continuous streaming permitting the exact advection of particles to their nearest lattice neighbours $ f_{\alpha}\left(\bm{x}+\bm{\xi}_{\alpha},t+\Delta t \right) $, thus ensuring that zero ‘numerical diffusion’ is generated \cite{Bernaschi_2019_RevModPhys.91.025004}. This intrinsic feature is a clear numerical advantage compared to conventional CFD solvers constructed on the basis of fractional advection (e.g., finite difference, finite volume, etc.) \cite{Krueger,Succi_2018}. 
This is particularly important, given that the discretisation of the advection term $\bm{u}\bcdot\bnabla\bm{u}$ in Eq.~(\ref{eq:NSE}) is generally the major source for numerical diffusion in conventional CFD solvers \cite{EKATERINARIS2005192}. It is worth highlighting that LBM has previously been shown to compare relatively well with spectral solvers (i.e., numerical solvers devoid of numerical diffusion) \cite{Spectral_POF,PENG2010568}.

Keeping with the inherent simplicity of LBM, whereas traditional CFD solvers possess a non-trivial pressure-velocity coupling, which requires solving an additional pressure-correcting Poisson equation iteratively (i.e., SIMPLE, PISO, etc.) to enforce continuity, the mesoscopic framework of LBM offers a much simpler and direct approach \cite{Krueger,NOURGALIEV2003117}. 
Pressure is instead implicitly captured through the density field $P\propto\rho$, avoiding the need to solve an additional Poisson equation iteratively. The simplicity of LBM also extends to its treatment of boundary conditions, which are implemented locally at the lattice nodes without requiring global matrix operations. This makes it particularly well-suited for complex geometries, specifically porous media \cite{HE2019160,Liu2014MultiphaseLB,PRE_POROUS,FATTAHI2016247} and moving boundaries \cite{lallemand2003lattice, strack2007three, owen2011efficient,huang2022streamline,huang2023power}. Additionally, its numerical formulation ensures straightforward enforcement of conditions like no-slip walls (e.g., bounce-back schemes), wherein distributions $f_{\alpha}$ are simply reflected back \cite{Krueger}.

The inclusion of additional multiphysics into LBM is relatively straightforward. As described in Subsection~\ref{SS:Forces}, additional momentum contributions, arising from electric forces \cite{Medvedev,Jiachen_POF,ZHANG2025109059}, magnetic forces \cite{SHEIKHOLESLAMI2015273}, structural reaction forces \cite{huang2021transition,wang2024wall,huang2024self}, capillary forces (Subsection \ref{SS:Multiphase}) \cite{Li_Surface,Surface_Tens_LBM,XU2015261,Chris_2019,Chris_2020}, active matter \cite{Carenza2019,Marenduzzo}, viscoelasticity \cite{Gupta_Hybrid,DZANIC2022105280,DZANIC_FEEDBACK,Chiyu_PRE,ma2023effects}, etc. are easily incorporated through an additional forcing term or by modifying the equilibrium stress. This feature allows LBM to conveniently extend itself to account for additional non-ideal interactions in multiphase and multicomponent problems. In LBM, phase separations can instead be generated automatically from particle dynamics; thus, no special treatment is required to manipulate the interfaces \cite{LBM_MICRO}. 

It is important to note that, like any numerical solver, LBM is not devoid of limitations; a number of caveats concern its appropriate application to solve fluid flow problems. 
While the intrinsic features of LBM (i.e., explicit nature and structured grids) facilitate the implementation of complex boundary conditions (e.g., porous media, non-ideal interactions, etc.) \cite{Krueger,LBM_MICRO,CHANG2009940} and effective parallelization \cite{GPU_LBM,GPU_O,GPU_RIN,GPU_XIAN}, it raises certain accuracy and stability issues. For instance, the inherent time-dependence of LBM makes it less efficient and not particularly well-suited for simulating steady-state flow problems, where iterative or direct solvers based on traditional methods (e.g., finite volume or finite element) may converge more quickly to the desired solution without the need to evolve the system over time \cite{GELLER2006888}. Moreover, whereas the deterministic stability constraint in traditional CFD approaches is based on the Courant number $C=|\bm{u}| \Delta t/\Delta x\leq1$, the BGK SRT approximation in LBM (Eq.~\ref{eq:BGK_Collision}) inherently imposes the additional stability constraint $\tau/\Delta t>0.5$ (i.e., a non-negative kinematic viscosity). A relaxation time-dependent viscosity is particularly problematic in flow regimes dealing with high Re \cite{Turb_LBM}, as well as non-Newtonian problems wherein the viscosity becomes shear-dependent \cite{Gabbanelli}. On the other hand, in problems where viscosity dominates (i.e., $Re\ll1$), to physically achieve such strong viscous effects typically requires $\tau>1$ (i.e., under-relaxation of $f_{\alpha}\rightarrow f_{\alpha}^{eq}$), which in itself introduces discretisation errors in the form of numerical diffusion \cite{Krueger,Behrend}, as well as solid wall boundary discretisation errors which scale with $\tau$ \cite{LBM_BC_3,Ginzburg}. Therefore, whereas hydrodynamic solutions in traditional CFD approaches are uniquely determined by their non-dimensional physical parameters, the standard BGK LBM exhibits $\tau$-dependent and therefore viscosity-dependent solutions. In general, these numerical issues can be alleviated through an appropriate selection of $0.5<\tau\leq1$ or through more advanced collision operators, such as TRT or MRT, which although permit viscosity-independent numerical solutions \cite{Humieres,Lallemand_MRT,Ginzburg}, still require the careful and often "problem-specific" tuning of multiple relaxation parameters. 

Although LBM is generally considered a computationally efficient CFD tool, challenges can indeed arise due to its inherently memory-intensive approach. Unlike traditional CFD methods, which primarily store macroscopic variables (e.g., velocity and pressure), LBM requires storing multiple distribution functions at each lattice site. Although this computational burden is often balanced out by its simple and highly parallelizable algorithm process, the memory requirements become increasingly problematic when considering large-scale (i.e., 3D) problems \cite{BANARI20141819,LBM3D_PROBLEM}, solving higher-order LBMs \cite{shan_2006,Chikatamarla}, or solving multiple LBM distribution functions to consider additional multiphysics \cite{Shan1995MulticomponentLM,Shan_Chen_1993,MALASPINAS20101637,SU_PRE,He2000_Heat,Chen_Heat_2LBM}. However, recent work has focused on the development of memory-efficient LBM schemes \cite{Memory_LBM,Velro,MATYKA2021108044}, which help mitigate the memory bottleneck and make LBM more feasible for large-scale and multiphysics simulations.

\section{Generalised Newtonian Fluids}
\label{sec:GNF}
\begin{figure*}[t]
    \centering
    \includegraphics[width=5.0in]{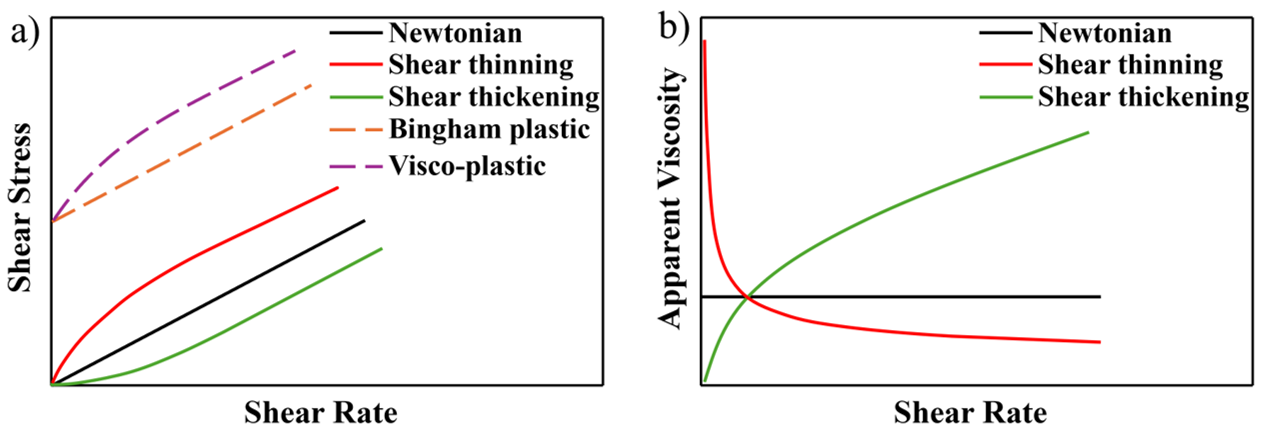}
    \caption{Qualitative flow curves illustrating the rheological behavior of various generalized Newtonian fluids (GNFs): a) shear stress ($\sigma_s$) as a function of shear rate ($\dot{\gamma}$), and b) apparent viscosity ($\mu$) as a function of shear rate ($\dot{\gamma}$). These curves demonstrate typical shear-thinning, shear-thickening, and viscoplastic responses characteristic of GNFs, highlighting the nonlinear dependence of stress and viscosity on the applied shear rate.}
    \label{fig:GFNs}
\end{figure*}
Generalised Newtonian fluids (GNFs) are distinguished from Newtonian fluids by their shear-dependent viscosity, which underpins a wide range of complex flow behaviors observed in both natural and engineered systems. Accurately capturing these rheological properties is essential for applications spanning from food processing to biomedical engineering, yet the diversity of responses exhibited by GNFs continues to present significant challenges for both experimental characterization and theoretical modeling \citep{bowers2021generalized}. This section reviews the fundamental physical principles and constitutive models that form the basis for understanding GNFs, providing a foundation for advanced computational approaches.

\subsection{Rheological characteristics and constitutive models}
Unlike Newtonian fluids, which exhibit a constant viscosity ($\mu$) regardless of the applied shear rate, GNFs are characterized by a viscosity (often termed apparent or effective viscosity) that is a nonlinear function of the local shear rate ($\dot{\gamma}$), but is independent of the material’s deformation history or elapsed time \citep{krishnan2010rheology,bowers2021generalized}. This means that the current value of the shear stress ($\sigma_s$) at a point in the fluid is solely determined by the instantaneous shear rate at that location, as illustrated in Figure~\ref{fig:GFNs}. GNFs, therefore, lack memory effects, and their constitutive behavior is adequately described by time-independent models. The most common non-Newtonian behaviors exhibited by GNFs include shear-thinning (pseudo-plastic), shear-thickening (dilatant), and visco-plastic responses, with or without shear-thinning characteristics \citep{krishnan2010rheology}.

The constitutive relationship for GNFs in simple shear flow is typically expressed as
\begin{equation}
\boldsymbol{\sigma}_s=2 \mu(\dot{\gamma}) \boldsymbol{D},
\end{equation}
where $\mu(\dot{\gamma})$ denotes the shear-rate-dependent viscosity function, and $\dot{\gamma}$ is the shear rate (a measure of the rate of deformation) defined by
\begin{equation}
\dot{\gamma}=\sqrt{2 \operatorname{tr}\left(\boldsymbol{D}^2\right)} = \sqrt{2 \boldsymbol{II}_{\boldsymbol{D}}},\label{sr}
\end{equation}
where $\boldsymbol{D}$ denotes the strain rate tensor, given by the symmetric component of the velocity gradient,
\begin{equation}
\boldsymbol{D}=\frac{1}{2}\left[\nabla \boldsymbol{u}+(\nabla \boldsymbol{u})^T\right],\label{srt}
\end{equation}
and $\boldsymbol{II}_{\boldsymbol{D}}$ is the second invariant of the strain rate tensor and is defined as the trace of the tensor $\boldsymbol{D}^2$, 
\begin{equation}
\boldsymbol{II}_{\boldsymbol{D}}=\boldsymbol{D}: \boldsymbol{D}=\operatorname{tr}(\boldsymbol{D}^2).
\end{equation}

The rheological diversity of GNFs is prominently displayed in their viscosity responses to shear rate ($\dot{\gamma}$), as illustrated qualitatively in Figure \ref{fig:GFNs} b). Shear-thinning fluids, exemplified by polymer melts and blood, exhibit a decrease in apparent viscosity ($\mu$) with increasing $\dot{\gamma}$. Conversely, shear-thickening fluids, such as dense cornstarch suspensions, show an increase in $\mu$ under rising $\dot{\gamma}$. To quantitatively capture these varied behaviors, a multitude of constitutive equations have been proposed in the literature. These models range significantly in complexity and theoretical foundation; some represent empirical formulations designed primarily to fit experimental viscometric data ($\sigma_s$ vs. $\dot{\gamma}$), while others possess a more rigorous theoretical basis derived from statistical mechanics or extensions of kinetic theory to liquids, albeit often blended with empirical adjustments \citep{krishnan2010rheology}. While comprehensive compilations of viscosity models have been presented in previous reviews by Phillips and Roberts \cite{phillips2011lattice} and, more recently, by Sun \textit{et al.} \cite{sun2024review}, we focus here on selected constitutive models that have gained widespread acceptance in rheological characterisation and numerical simulation.

The power-law model, also known as the Ostwald-de Waele model, is a generalized Newtonian constitutive equation widely employed to characterize shear-thinning and shear-thickening fluids, including blood and polymer solutions. For simple shear flow, the model relates shear stress ($\sigma_s$) to shear rate ($\dot{\gamma}$) as, 
\begin{equation}
\sigma_s=m\dot{\gamma}^n,
\end{equation}
where the apparent viscosity ($\mu$) becomes
\begin{equation}
\mu=m\dot{\gamma}^{n-1}. \label{plf}
\end{equation}
Here, $m$ represents the fluid consistency coefficient (units $Pa \cdot s^n$), providing a measure of the fluid's average viscosity, while $n$ is the dimensionless flow behavior index, characterizing the deviation from Newtonian behavior. Both $m$ and $n$ are empirical parameters, typically obtained by fitting the model to experimental rheological data over a specific range of shear rates. The value of $n$ dictates the nature of the fluid's response to shear. For $0<n<1$, the fluid exhibits shear-thinning (pseudo-plastic) behavior, where apparent viscosity decreases with increasing shear rate $(\mathrm{d} \mu / \mathrm{d} \dot{\gamma})<0$. The smaller the value of $n$, the more pronounced the shear-thinning effect. For $n=1$, the model reduces to the Newtonian case, where viscosity ($\mu = m$) is constant and independent of shear rate. For $n>1$, the fluid displays shear-thickening (dilatant) behavior, with apparent viscosity increasing with shear rate $(\mathrm{d} \mu / \mathrm{d} \dot{\gamma})>0$.

The power-law model finds extensive application due to its simplicity. It is frequently used to approximate the behavior of blood, where $n$ typically ranges from 0.6 to 0.9, influenced by factors such as hematocrit, fibrinogen, and cholesterol concentrations \citep{hussain1999relationship,shibeshi2005rheology}. Similarly, many polymer melts and solutions are adequately described by the power-law model within specific shear rate ranges, often with $n$ values between 0.3 and 0.7, depending on polymer concentration and molecular weight \citep{krishnan2010rheology}. Although the power-law model provides a simple and effective approximation for shear-thinning behavior, it is inherently limited in that it does not capture the Newtonian viscosity plateaus at either very low ($\dot{\gamma} \rightarrow 0$) or very high ($\dot{\gamma} \rightarrow \infty$) shear rates. In practice, the empirical parameters $m$ and $n$ remain approximately constant only within a restricted range of shear rates, necessitating prior knowledge of the operational shear rate regime for accurate model application. To address the unbounded viscosity predictions of the power-law model outside its valid range, numerical implementations often employ truncated forms, constraining the viscosity within prescribed lower and upper bounds ($\mu_0 < \mu < \mu_\infty$), corresponding to the asymptotic viscosities at low and high shear rates, respectively. This approach, while practical for numerical stability and convergence, does not resolve the underlying physical limitations of the model.

To address the limitations of the power-law model, Cross \citep{cross1965rheology} introduced a four-parameter constitutive equation capable of predicting both the zero-shear ($\mu_0$) and infinite-shear ($\mu_\infty$) viscosity plateaus. Building upon this foundation, Carreau \citep{carreau1972rheological} proposed a modified formulation that explicitly links polymer chain dynamics to macroscopic rheology. This model has been widely used for modelling shear-thinning fluids. To better characterise the low-shear-rate behaviour, Yasuda \citep{yasuda1979investigation} extended the Carreau model by incorporating an extra parameter, which enables more accurate fitting of experimental data for fluids. For concentrated disperse systems such as blood, the Quemada model \citep{quemada1978rheology} provides a unique framework that integrates shear rate and hematocrit dependence. This model is particularly effective for predicting the shear-thinning behavior of blood across physiologically relevant hematocrit levels (30-50\%).

Recent advances in modelling shear-thickening fluids propose a structural framework based on the Effective Volume Fraction (EVF) concept \citep{quemada1998rheological,mari2015discontinuous}. In concentrated suspensions, particles form transient structural units (SUs)-compact clusters or primary particles-that aggregate into larger networks at low shear rates. As $\dot{\gamma}$ increases, hydrodynamic forces disrupt these networks, initially causing shear thinning \citep{athani2025transients}. Beyond a critical shear rate, hydrodynamic clustering dominates, increasing the EVF and triggering dilatancy \citep{ovarlez2020density}. This transition arises from the competition between hydrodynamic forces (promote the formation of shear-resistant clusters) and Brownian motion (favors particle dispersion at lower $\dot{\gamma}$) \citep{lee2022microstructure,liu2023shear}.

Viscoplastic fluids exhibit a yield stress ($\sigma_0$), a critical threshold that must be exceeded for flow to occur. Below $\sigma_0$, the material behaves as an elastic solid or moves rigidly; above $\sigma_0$, it flows as a fluid, often exhibiting Newtonian or shear-thinning behaviour. This dual solid-fluid response is ubiquitous in industrial and biological systems, including blood, cement suspensions, polymer drilling fluids, and food products such as yogurt, molten chocolate, and tomato puree. Comprehensive reviews of viscoplastic rheology and fluid mechanics are are available in the literature \citep{bird1983rheology,barnes1999yield,sun2024review}. A range of non-Newtonian constitutive models has been developed to describe the complex flow behaviour of visco-plastic fluids. Among the most widely used are the Bingham, Herschel-Bulkley, and Casson models, as well as their respective modifications and extensions, such as the modified Casson, Casson type K-L, Heinz-Casson, Mizrahi-Berk, Vipulanandan, and Robertson-Stiff models. The Bingham plastic model represents the simplest viscoplastic formulation, characterised by a linear relationship between shear stress and shear rate beyond the yield point. As illustrated in Figure \ref{fig:GFNs} a), the constitutive equation for one-dimensional shear is expressed as, 
\begin{align*}
\sigma_s &=\sigma_0+\mu_p \dot{\gamma},  \quad \sigma_s>\sigma_0 \\
\dot{\gamma} &=0, \quad \quad \quad \quad \sigma_s \leqslant \sigma_0,  
\end{align*}
where $\mu_p$ denotes the plastic viscosity, representing the slope of the flow curve above the yield stress. The corresponding shear rate ($\dot{\gamma}$) can be derived as,
\begin{equation}
\dot{\gamma}= \begin{cases}\frac{\sigma_s-\sigma_0}{\mu_p}, & \sigma_s>\sigma_0 \\ 0, & \sigma_s \leqslant \sigma_0\end{cases}.
\end{equation}
The apparent viscosity, defined as the ratio of shear stress to shear rate, is given by
\begin{equation}
\mu= \begin{cases}\frac{\sigma_0}{\dot{\gamma}}+\mu_p, & \sigma_s>\sigma_0 \\ \infty, & \sigma_s \leqslant \sigma_0\end{cases}.
\end{equation}

The Bingham model provides a foundational framework for describing viscoplastic materials such as industrial suspensions (cement, drilling muds), coatings (paints), and biological fluids (blood in capillaries) \citep{zhu2005non,dean2007numerical}. Owing to its simple constitutive, the model makes it easy to determine the yield stress $\sigma_0$, which is the critical threshold at which the material transitions from solid-like to fluid-like behavior. However, the model fails to capture nonlinear rheological responses due to its linear post-yield stress-shear rate relationship \citep{khalil2011rheological}, which is improved by the Herschel-Bulkley model that introduces a power-law dependence on shear rate for stress exceeding the yield point. For one-dimensional shear flow, the Herschel-Bulkley model \citep{herschel1926konsistenzmessungen} is expressed as,
\begin{equation}
\begin{array}{ll}
\sigma_s=\sigma_0+m \dot{\gamma}^n, & \sigma_s>\sigma_0 \\
\dot{\gamma}=0, & \sigma_s \leqslant \sigma_0
\end{array},
\end{equation}
with apparent viscosity defined by 
\begin{equation}
\mu= \begin{cases}\frac{\sigma_0}{\dot{\gamma}}+m \dot{\gamma}^{n-1}, & \sigma_s>\sigma_0 \\ \infty, & \sigma_s \leqslant \sigma_0\end{cases}.
\end{equation}
The first term ($\sigma_0/\dot{\gamma}$) dominates at low shear rates, while the second term ($m\dot{\gamma}^{n-1}$) governs shear-thinning or thickening behavior at higher $\dot{\gamma}$.

Numerous studies have demonstrated that the rheological behaviour of polymer drilling fluids and cement slurries is well described by the Herschel-Bulkley model, which effectively captures both yield stress and post-yield shear-thinning characteristics \citep{kelessidis2006optimal,kannojiya2020simulation}. In addition to the Bingham and Herschel-Bulkley models, the Casson model is widely employed for viscoplastic fluids, particularly in applications involving complex suspensions and biological fluids \citep{walawender1975approximate}. The Casson model, which can be viewed as a modification of the Bingham model with all terms raised to the half power, was originally developed to relate the viscosity of a suspension to changes in its microstructure, conceptualizing the dispersed phase as elongated cylindrical aggregates with high aspect ratios. The Casson model is particularly suitable for describing the rheology of fluids such as chocolate and blood at low shear rates and for blood samples with red blood cell volume fractions below 40\% \citep{phillips2011lattice}. To further enhance its predictive capability, several extensions have been developed. To capture the non-linear viscosity response, the modified Casson model incorporates strain-rate correction terms, and the K-L (Krieger-Lodge) model incorporates both yield stress and a transition from shear-thinning to a high-shear Newtonian plateau \citep{papanastasiou1987flows,luo1992study}. In simulations of blood flow, the modified Casson model typically predicts higher viscosities under Newtonian conditions and lower viscosities at low shear rates compared to the original Casson formulation \citep{razavi2011numerical}. Specifically designed to describe human blood, the K-L model incorporates a yield stress, a pronounced shear-thinning region, and a transition toward a Newtonian plateau at high shear rates. This model has demonstrated its effectiveness in capturing the shear-thinning response of blood that outperforms the Casson and Newtonian models.

These advanced constitutive models provide a comprehensive framework for modelling the complex rheological behaviour of viscoplastic fluids in a wide range of industrial and biomedical applications. However, their limitations need to be paid attention to. For example, GNF models neglect normal stress differences or elastic effects present in viscoelastic fluids. Furthermore, GNFs respond instantaneously to changes in shear rate, neglecting memory effects and time-dependent relaxation phenomena. Finally, these models are generally unsuitable for predicting fluid behaviour in extensional or elongational flows, such as those encountered in fibre spinning or certain polymer processing operations. As such, careful model selection and validation against experimental data remain essential for accurate simulation and analysis of non-Newtonian fluid systems.

\subsection{LBMs for generalised Newtonian fluids simulations}
The adaptation of LBM for GNFs involves incorporating a shear-rate-dependent viscosity into the collision operator. In the standard Newtonian LBM, the kinematic viscosity relates to the relaxation time through,
\begin{equation}
\nu=c_s^2\left(\tau-\frac{1}{2}\right) \frac{\Delta x^2}{\Delta t}.
\end{equation}
For GNFs, this relationship becomes spatially and temporally varying,
\begin{equation}
\nu(\boldsymbol{x}, t)=\frac{\mu(\dot{\gamma}(\boldsymbol{x}, t))}{\rho}=c_s^2\left(\tau(\boldsymbol{x}, t)-\frac{1}{2}\right) \frac{\Delta x^2}{\Delta t},
\end{equation}
where the apparent viscosity $\mu(\dot{\gamma})$ depends on the local shear rate $\dot{\gamma}$.

The shear rate $\dot{\gamma}$ (Eq. \ref{sr}) and strain rate tensor $\boldsymbol{D}$ (Eq. \ref{srt}) can be calculated at each point on the lattice by finite-difference formulate applied to the nearby grid point velocity $\boldsymbol{u}$ \citep{gabbanelli2005lattice,huang2021low}. Alternatively, the local shear rate and strain rate tensor can be computed directly from the non-equilibrium distribution functions ($f_\alpha^{neq}=f_\alpha-f_\alpha^{e q}$) without requiring explicit velocity gradient calculations \citep{phillips2011lattice},
\begin{equation}
\dot{\gamma}=\sqrt{2 \operatorname{tr}\left(\boldsymbol{D}^2\right)}, \quad \boldsymbol{D}_{i j}=-\frac{1}{2 \rho \tau c_s^2} \sum_{\alpha} \boldsymbol{\xi}_{\alpha i} \boldsymbol{\xi}_{\alpha j} \left(f_\alpha-f_\alpha^{e q}\right). 
\end{equation}
This computation is entirely local, eliminating the need for finite difference approximations of velocity gradients, thereby reducing numerical diffusion and retaining full second-order accuracy \citep{kruger2009shear}. This local method has been adopted by most of the recent LBM studies on GNFs \citep{conrad2015accuracy,hill2024development}. The disadvantage arises because $\dot{\gamma}$ is a function of $\tau$, resulting in the following implicit expression, 
\begin{equation}
\tau=\frac{\mu(\dot{\gamma}(\tau)) \Delta t}{\rho c_{s}^2 \Delta x^2}+\frac{1}{2}.\label{tau2}
\end{equation}
Eq. \ref{tau2} can be solved numerically at each grid point using an iterative method, which can sometimes be computationally expensive. As an alternative, $\dot{\gamma}(\tau)$ can be approximated by using the value of $\tau$ from the previous time step, which provides an explicit and efficient calculation for $\tau$ \citep{phillips2011lattice}. 

The implementation of various GNF constitutive models within the LBM framework requires systematic evaluation of numerical accuracy and stability. For power-law models, the apparent viscosity is given by Eq. \ref{plf}, thereby producing a shear-rate-dependent relaxation time,  
\begin{equation}
\tau(\dot{\gamma})=\frac{m \dot{\gamma}^{n-1}\Delta t}{\rho c_s^2 \Delta x^2}+\frac{1}{2}.
\end{equation}
For improved numerical stability and convergence, truncated forms that constrain the viscosity within prescribed lower and upper limits are often adopted \citep{gabbanelli2005lattice}. For yield-stress fluids described by the Bingham or Herschel-Bulkley models, the implementation becomes more complex due to the discontinuity in their viscosity functions. Specifically, for $\sigma_s>\sigma_0$ in the Herschel-Bulkley model, the relaxation time can become extremely large as $\dot{\gamma} \to 0$, potentially causing numerical instabilities. One effective way to address this issue is the regularization method \citep{ginzburg2002free}, where Papanastasiou-type regularization \citep{papanastasiou1987flows} smooths the discontinuity by introducing an exponential function,
\begin{equation}
\mu(\dot{\gamma})=\left(\frac{\sigma_0}{\dot{\gamma}}[1-\exp (-\alpha \dot{\gamma})]+m \dot{\gamma}^{n-1}\right),
\end{equation}
where $\alpha$ has dimensions of time and is a regularization parameter (exponential growth factor) that controls the transition smoothness of the viscosity \citep{burgos1999determination}.

\subsection{Applications and challenges}
This section reviews major methodological advances, validation studies, practical applications, computational challenges, and recent trends in the application of the LBM to GNFs. Since the foundational work of Aharonov and Rothman \citep{aharonov1993non} on simulating non-Newtonian flow in porous media by incorporating shear-dependent viscosity through collision operator modifications, this field has seen significant advancement. Building on this foundation, the seminal study by Gabbanelli \citep{gabbanelli2005lattice} introduced a truncated power-law model for shear thinning and shear thickening fluids and demonstrated that the \textit{ad hoc} modification of the LBM accurately describes the flow. Yoshino \textit{et al.} \citep{yoshino2007numerical} introduced a new numerical method that calculates shear-dependent viscosity via a variable parameter linked to local shear rate, and maintaining a unity relaxation time for enhanced numerical stability. This method has been shown to yield greater accuracy than the standard approach of making the relaxation time a function of local shear rate, for modelling both shear-thickening and shear-thinning behaviors in 2D reentrant corners and 3D porous structures. Chai \textit{et al.} \citep{chai2011multiple} presented an MRT LBM model for calculating the strain rate tensor in GNFs, and validated the model with the accuracy and stability being studied. Leonardi \textit{et al.} \citep{leonardi2011numerical} presented a new approach for numerical rheometry of bulk materials using non-Newtonian LBM coupled with discrete element methods (DEM). In their 2D and 3D validation simulations, the power-law model was found to be able to reproduce the characteristics of shear-thinning and shear-thickening behaviour, while the regularised Bhingham model was found to be prohibitively unstable. Conrad \textit{et al.} \citep{conrad2015accuracy} provided a comprehensive accuracy analysis by introducing a dimensionless collision frequency dependent error for power-law fluids with both SRT and MRT non-Newtonian LBM model, revealing that the simulation is about one order of magnitude more accurate for an optimal simulation Mach number ($Ma$). They found that numerical diffusion error and compressibility errors were the dominant sources of error, and an expression for deducing a best choice for the $Ma$ was derived. While more accurate, the free relaxation parameters of the MRT model have been shown to have a significantly impact on the error in velocity and the optimal $Ma$. Although the SRT model is computationally efficient, it fails to maintain second-order convergence. Adam \textit{et al.} \citep{adam2021cascaded} presented a new 3D cascaded LBM for power-law fluids, demonstrating improved accuracy, second-order convergence, and significant improvements in numerical stability over the SRT and standard MRT LBM models. The most recent methodological development is the enhanced double distribution function method LBM that independently models the momentum and scalar variables, such as temperature or concentration \citep{vaseghnia2025enhanced}.

Additional comprehensive validation studies have been conducted in this field. Malaspinas \textit{et al.} \citep{malaspinas2007simulation} demonstrated second-order spatial accuracy for a power-law model in Poiseuille flow through comparison with analytical solutions. Their comparative study of 4:1 planar contractions, using both power-law and Carreau-law models, showed excellent agreement with commercial finite element solver predictions. Bisht \textit{et al.} \citep{bisht2021assessment} presented a benchmark study of MRT LBM for multiple non-Newtonian constitutive models, including Power-law, Carreau, Carreau-Yasuda, and Cross, using four standard 2D benchmark flow configurations, and reported results agree well with published data. The accuracy of LBM for various non-Newtonian flows has been validated by integrating thermal effects \citep{wang2016regularized,wang2019effects}, particle dynamics \citep{leonardi2011numerical,wang2019effects}, fluid-structure interactions \citep{huang2021low}, complex geometries \citep{hill2021lattice,daeian2025multi}, multiphase flow \citep{hill2024development}, and GPU accelerated MRT-LBM \citep{molla2025three}. The collective validation findings demonstrate LBM as a robust, accurate, and computationally efficient framework for GNF simulations, especially outperforming alternatives in complex geometries, multi-physics applications, and scalable parallel computing architectures.

The versatility and success of LBM in GNF simulations have been evidenced by its diverse applications. Leonardi et al. \citep{leonardi2011numerical} characterized bulk material flows using non-Newtonian LBM coupled with DEM. By implementing power-law and Bingham models in a cylindrical Couette rheometer, they showcase the method’s capability for numerical rheometry of complex materials such as Leighton Buzzard sand and synthetic soil samples. Recent advancements have significantly expanded the capability to incorporate thermal effects in non-Newtonian flows. Dong et al. \citep{dong2018numerical} implemented thermal LBM on non-orthogonal grids and couple with a power-law model by using double-distribution-function method, showing accurate simulation of power-law fluids in irregular geometries with forced and natural convection. He et al. \citep{he2024numerical} developed a phase-field-based LB model for thermocapillary motion of two-phase non-Newtonian power-law fluids, three evolution equations were coupled and solved to capture phase interface variation, power-law fluid dynamics, and temperature, respectively. For the interactions between particles and non-Newtonian fluids, Delouei et al. \citep{delouei2016non} investigated particle interactions in power-law fluids in the computational framework of the immersed boundary-LBM, showing significant differences in particle sedimentation such as kissing time and transverse position of particle during tumbling intervals. Using similar computational framework, Jiao et al. \citep{jiao2022numerical} investigated different shapes of particle sedimentation in power-law fluids, revealing that shear-thickening fluids has a longer kissing duration than the Newtonian and shear-thinning fluids. For the use of non-Newtonian LBM in simulating blood flow, Bernsdorf and Wang \cite{bernsdorf2009non} simulated blood flow in a patient-specific cerebral aneurysm using SRT LBM with the Carreau-Yasuda model for capturing the shear-thinning effect of blood, showing a lower viscosity and wall shear stress when compared to Newtonian flow. Hill and Leonardi \citep{hill2021lattice} presented a TRT LBM with Kuang-Luo rheological model to capture the viscoplasticity and pseudoplasticity of blood, and validation was conducted against analytical solutions with the application of blood flow in a carotid artery.

Despite its advantages, LBM for GNFs faces several inherent limitations that must be carefully considered. A central numerical challenge in non-Newtonian LBM is the strong dependence of the collision relaxation time ($\tau$) on the local viscosity. For GNFs, $\tau$ varies throughout the domain according to spatial variations in viscosity. When $\tau \to 0.5$ (corresponding to low viscosity), simulations become susceptible to numerical instability. While advanced collision models, such as MRT, provide enhanced numerical stability, they typically require fine-tuning of parameters and can be challenging to optimise \citep{conrad2015accuracy,bisht2021assessment}. The cascade LBM based on central moments \citep{adam2021cascaded} shows enhanced numerical stability over SRT and MRT collision operators. When $\tau \gg 1$ (high viscosity), loss of accuracy and convergence issues may occur, as seen in LBM studies for Herschel-Bulkley and Bingham fluids. Conrad and Bösch \citep{conrad2015accuracy} demonstrated that the choice of simulation Mach number and grid resolution directly influences collision frequency-dependent errors. While fine-tuning these parameters can improve results (sometimes by an order of magnitude), second-order accuracy cannot always be achieved for every parameter set. This sensitivity is especially problematic for flows with rapidly evolving or extreme shear rate distributions. Vaseghnia et al. \citep{vaseghnia2024evaluation} showed that the wall shear stress distribution accuracy and the numerical stability can be significantly improved with carefully tuned relaxation times in the MRT collision operator. 

Recent advances in grid refinement have made the application of LBM to GNFs more efficient and accurate, marking a key trend in the field. Daeian et al. \citep{daeian2025multi} developed a multi-domain grid refinement method for MRT LBM, specifically targeting non-Newtonian blood flow within complex vascular geometries. To mitigate the heavy memory and computational demands of large-scale non-Newtonian LBM simulations, Chen et al. \citep{chen2020simplified} proposed a simplified LBM scheme that evolves macroscopic variables directly rather than the standard distribution functions, expecting that 33.3\% of virtual memory can be reduced for a 2D model compared with conventional LBM. LBM is increasingly being utilised to investigate intricate flow phenomena in engineering and physical sciences. Yu et al. \citep{yu2025lattice} demonstrated how rheology-dependent effects, such as negative Magnus forces and altered wake interference patterns, emerge in power-law fluid flow past tandem cylinders. The inherent parallel nature of LBM has driven the development of advanced implementations for high-performance computing frameworks. Molla et al. \citep{molla2025three} showcased GPU-accelerated MRT-LBM computation using CUDA, enabling stable and efficient modelling of Herschel-Bulkley fluids in a 3D lid-driven cubic cavity. Future work may leverage data-driven approaches for parameter optimization and automated prediction of relaxation time in MRT collision operators, adapted to local flow conditions. Such strategies offer a promising pathway for further improving both simulation accuracy and robustness, especially for complex and multi-physics problems. Further development of specialised collision operators that better handle extreme viscosity variations while ensuring numerical stability remains a highly desirable direction.

\begin{figure*}[t]
	\centering
	\includegraphics[width=0.9\textwidth]{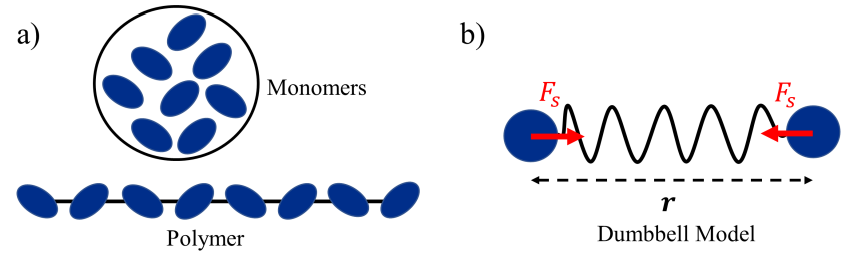}
	\caption{Illustration of a) the molecular build-up of long-chain polymers comprised of
binding monomers, which can b) be simplified towards a coarse-grained treatment, known as the
dumbbell model. The dumbbell model consists of two beads connected by a spring of distance $\bm{r}$, representing the entropic elasticity of the polymer chain. The beads account for the hydrodynamic drag experienced by the polymer, while the spring captures the force dipole $\bm{F}_s$ arising from the entropic restoring force, providing a mesoscopic description of polymer dynamics in a solvent }
	\label{figure_4}
\end{figure*}
\section{Viscoelastic Flows} \label{sec:Viscoelastic}
\noindent In addition to a shear-dependent viscous response, non-Newtonian fluids can also exhibit elasticity, characterised by the build-up of normal stress differences and time-dependent responses, such as stress relaxation and creep. This section of the review will first focus on the physical description of viscoelastic fluids followed by how the hydrodynamic LB solver has been creatively altered
in various attempts to include these so-called viscoelastic effects either macroscopically or in the
spirit of the LB method, mesoscopically.

Viscoelastic fluids, typically referring to the addition of polymers to a solvent, give rise to an additional nonlinear material property in the form of an elastic polymer stress response $\bm{\sigma}_p$, such that the total stress becomes $\bm{\sigma}=\bm{\sigma}_s+\bm{\sigma}_p$. However, whereas
the understanding of $\bm{\sigma}_s$ is relatively straightforward (see Section.~\ref{sec:GNF}), physically describing the microstructural behaviour of polymers through macroscopic quantities, such as $\bm{\sigma}_p$ , is far less
trivial and requires certain assumptions and simplifications.

From a molecular framework, the additional polymers in viscoelastic fluids can be described as large molecules made up of long chains of smaller molecules called monomers, as illustrated in Fig.~\ref{figure_4} a). To simplify the degrees of freedom, these polymer macromolecules are instead generally coarse-grained as an elastic dumbbell \cite{Oldroyd1950OnTF,Bird_Book,bird1983rheology}. This dumbbell model approach \cite{Oldroyd1950OnTF,Bird_Book,W_Kuhn} assumes a dilute
polymer solution and simply consists of treating polymer molecules as
two beads, representing  blocks of monomer, connected by a Hookean spring with a separation vector $\bm{r}$, as shown in Fig.~\ref{figure_4} b).
Notably, the dumbbell model is based on kinetic theory and captures the fundamental behaviour
of polymer molecules, as it permits relative stretching and relaxation of polymers in reference to
their initial equilibrium state. The separation between beads is essentially driven by a range of
interactions, including the viscous forces by the surrounding solvent fluid, the connecting spring
force, as well as Brownian forces \cite{Alves_Rev,Squires_2005}. After stretching,
the polymer experiences a relaxation back towards equilibrium, whereby the two beads are pulled
toward the spring, each exerting a force $\bm{F}_s$ back on the fluid, resulting in a force-dipole flow
and an anisotropic contribution to the normal stress. 

The dumbbell model provides a microscopic foundation for several macroscopic constitutive equations used to describe viscoelastic fluids, which can be written in a general form based on the total stress contribution $\bm{\sigma}$,
\begin{equation} \label{Eq:General_Polymer}
\bm{\sigma} + \lambda\mathscr{D}_t\bm{\sigma} = 2\mu_t\left[\bm{S}+\lambda_r\mathscr{D}_t\bm{S}\right],
\end{equation}
where $\lambda$ is the characteristic polymer relaxation time, $\mu_t\equiv\mu_s+\mu_p$ is the total zero-shear-rate viscosity of the solution, and $\lambda_r\equiv\lambda\mu_s/\mu_t$ is the retardation time. The operator $\mathscr{D}_t\bm{\sigma}$ represents a time derivative, which under the assumptions of low polymer deformation, the linearised Maxwell and Jeffrey's model both take $\mathscr{D}_t\bm{\sigma}=\partial_t\bm{\sigma}$ \cite{Bird_Book}. However, in the case of larger polymer deformation, characterised by moderate to high Weissenberg numbers $Wi=\lambda\dot{\gamma}\gg0$, the simple time derivatives in linear models are not frame invariant \cite{Bird_Book,HINCH2021104668}, and hence, require extension towards upper-convected $\mathscr{D}_t\bm{\sigma}=\overset{\nabla}{\bm{\sigma}}=\partial_t\bm{\sigma}+\bm{u}\bcdot\nabla\bm{\sigma}-\bm{\sigma}\bcdot\bnabla \bm{u}-(\bnabla\bm{u})^T\bcdot\bm{\sigma}$ or lower-convected $\mathscr{D}_t\bm{\sigma}=\overset{\Delta}{\bm{\sigma}}=\partial_t\bm{\sigma}+u\bcdot\nabla\bm{\sigma}+\bm{\sigma}\bcdot(\bnabla \bm{u})^T+(\bnabla\bm{u})\bcdot\bm{\sigma}$ time derivatives to maintain material objectivity of $\bm{\sigma}$ when translating, rotating, and deforming with the
local flow \cite{Oldroyd1950OnTF,Howard_Stone_Convected}. However, given the rod-like nature of polymers, the upper-convected derivative $\mathscr{D}_t\bm{\sigma}=\overset{\nabla}{\bm{\sigma}}$ is often preferred, giving rise to the popular Oldroyd-B model \cite{Oldroyd1950OnTF}, as it better captures the stretching and alignment of polymer chains with the flow (i.e., contra-variant nature), leading to more accurate predictions of positive normal stress differences and other nonlinear viscoelastic phenomena \cite{HINCH2021104668,RENARDY2021104573}.

The evolution of the polymer stress $\bm{\sigma}_p$ is generally expressed and numerically solved for in terms of a geometric second-order tensor $\bm{C}$, commonly referred to as the conformation tensor \cite{Alves_Rev}. Specifically, $\bm{C}$, is a measure of the second-order moment of the normalised end-to-end distance vector
of the polymer dumbbell, $\bm{C}\equiv\langle r_i r_j\rangle/r_{eq}^2$, where $\bm{r}_{eq}$ is the end-to-end vector magnitude at equilibrium, so that $\bm{C} = \bm{I}$ at rest \cite{Alves_Rev,VAITHIANATHAN20031}. From this definition, it follows that the conformation tensor is inherently a symmetric positive definite (SPD) matrix. Thus, in the dilute regime (i.e., $\lambda_r=\lambda$), the evolution for $\bm{C}$ becomes,
 \begin{equation}\label{eq:C_Eq}
    \frac{\partial \bm{C}}{\partial t}
 + \bm{u} \cdot \bnabla\bm{C} = \bm{C}\bcdot\left(\bnabla\bm{u}\right)+\left(\bnabla\bm{u}\right)^{{T}} \bcdot \bm{C}-\frac{f}{\lambda}\left(\bm{C}-\bm{I}\right),
\end{equation}
which is easily related back to the polymer elastic stress tensor, $\bm{\sigma}_p=\frac{\mu_p}{\lambda}\left(f\bm{C}-\bm{I}\right)$. Here, $f$ is a model parameter that allows different constitutive polymer models to be represented, most notably, the purely elastic Oldroyd-B ($f=1$) \cite{Oldroyd1950OnTF}, or models with finite extensibility like FENE-P ($f=(L^2-\mathrm{tr}\bm{I})/(L^2-\mathrm{tr}\bm{C})$) \cite{PETERLIN1961257}, where $L^2$ is the maximum polymer extension, and the linear Phan-Thien–Tanner (PTT) model ($f=1+\varepsilon(\mathrm{tr}\bm{C}-\mathrm{tr}\bm{I})$ for concentrated polymer solutions \cite{THIEN1977353,THIEN2}, with $\varepsilon$ controlling the degree of shear thinning.

(i.e., $f=1$ simplifies to the Oldroyd-B model) \cite{PETERLIN1961257,Bird_Book,Alves_Rev}. This additional polymer feedback contribution is easily incorporated into the hydrodynamic field through an additional momentum contribution,
\begin{equation}\label{eq:NSE_VISCO}
    \bnabla \bcdot \bm{u} = 0, ~~~ \rho\left(\frac{\partial \bm{u}}{\partial t} + \bm{u} \cdot \bnabla \bm{u}\right) = - \nabla p + \mu_s \bnabla^2 \bm{u} + \bnabla\cdot\bm{\sigma}_p.
\end{equation}
Multiple excellent reviews have already discussed how to solve Eqs.~\ref{eq:C_Eq} and \ref{eq:NSE_VISCO} based on conventional CFD techniques, such as finite difference schemes \cite{Crochet1983NumericalMI,Crochet_REV}, finite element methods \cite{PTBAAIJENS1998361}, and finite volume methods \cite{Alves_Rev}. The application of mesoscopic approaches, specifically LBM, to numerically simulate viscoelastic fluids was reviewed over a decade ago by Phillips \& Roberts \cite{Phillips_2011}, with a small section dedicated towards the end of the more recent review by Alves, Oliviera, \& Pinho \cite{Alves_Rev}. In what follows, we thoroughly document the most notable attempts made at incorporating viscoelasticity through LBM, as well as the recent exciting applications.

\subsection{MRT models}
\begin{figure*}[t]
	\centering
	\includegraphics[width=0.94\textwidth]{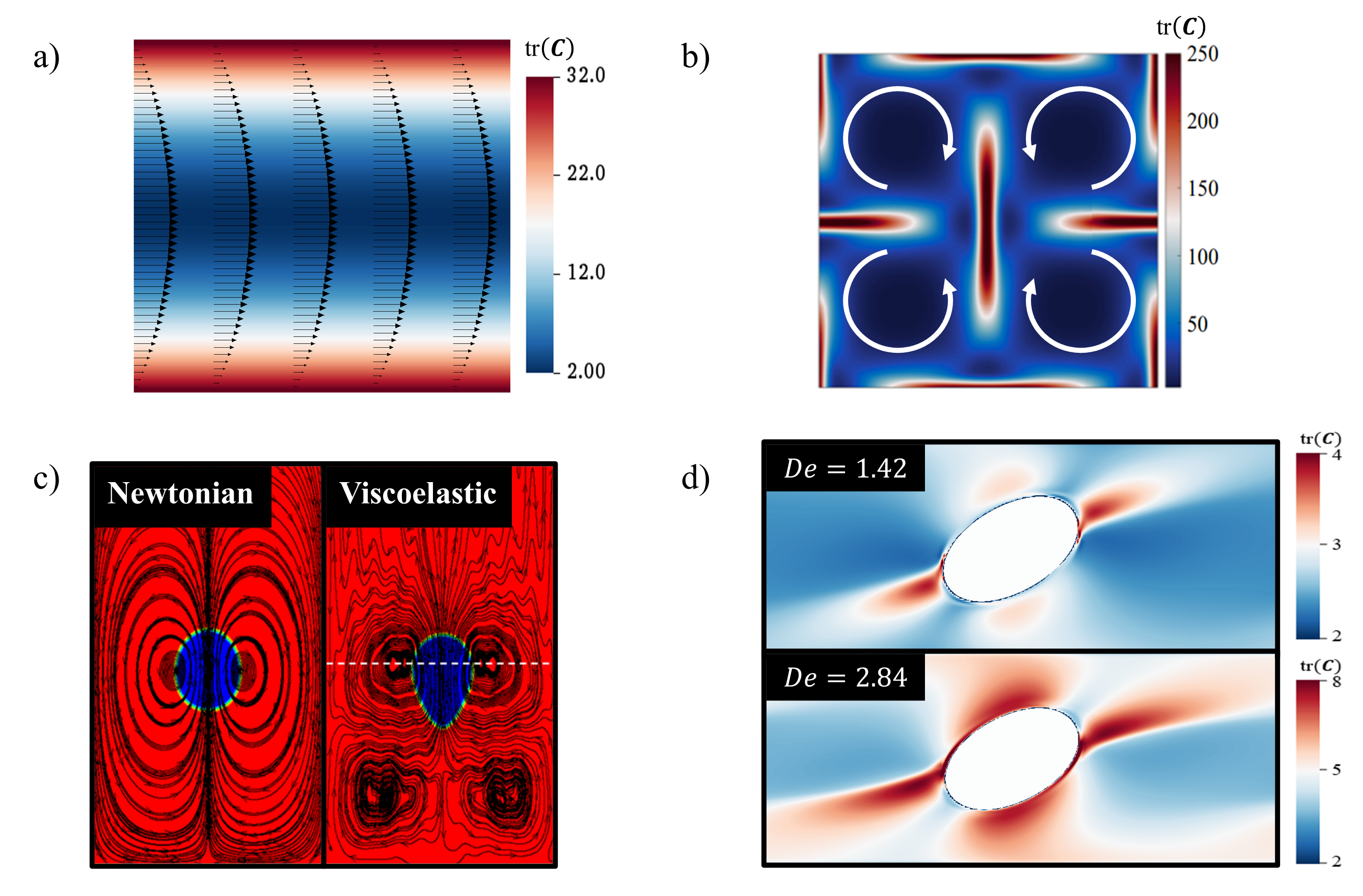}
	\caption{Several single-and-multiphase viscoelastic benchmark cases have emerged in validating the numerical performance of viscoelastic LBM solvers. a) The steady and time-dependent planar Poiseuille flow cases remain a useful starting point for benchmarking the numerical accuracy of LBM solvers with the well-known analytical solutions for the hydrodynamic and polymer field variables, as done in previous studies \cite{DZANIC_FEEDBACK,MALASPINAS20101637,SU_PRE,OSMANLIC2016190,Kuron,ZHANG2025105351}. b) A more stringent benchmark case is the four-roll mill problem, originally proposed by Thomases \& Shelley \cite{T_S_2007}, which imposes a constant external forcing that generates four rotating and counter-rotating vortices. In doing so, a strong extensional flow regime is defined with central stagnation points that help stretch polymers significantly. An analytical prediction for the steady-state stress in the vicinity of the central stagnation points exists \cite{T_S_2007}, which has been used to validate the numerical accuracy of LBM solvers \cite{DZANIC2022105280,DZANIC_FEEDBACK, MALASPINAS20101637, Kuron}. The numerical robustness of different viscoelastic LBM models have also been tested using the four-roll mill at higher elastic effects, with a useful stability condition based on the polymer field provided by Dzanic \textit{et al.} \cite{DZANIC_FEEDBACK}. c) and d) The popularity of multiphase LBM solvers have naturally extended themselves to viscoelastic problems, specifically, for the qualitative validation of a Newtonian droplet rising in a viscoelastic medium \cite{Chiyu_PRE,WAGNER2000227,Wagner_2005,Di_ADE_2019,YOSHINO}, as well as a Newtonian (or viscoelastic) droplet suspended in a viscoelastic (or Newtonian) medium under shear flow \cite{Gupta_Hybrid,Di_ADE_2019,YOSHINO,Dzanic_POF_Mobilization,Gupta_PRF,Gupta_Euro,Jiachen_POF}. Image in c) retrieved from Xie et al. \cite{Chiyu_PRE}.         }
	\label{figure_5}
\end{figure*}

\noindent Perhaps the earliest attempt made at coupling the characteristic behaviour of viscoelastic
fluids to LBM, involved altering the collision process through modified MRT schemes. This was initially attempted by Giraud \textit{et al.} \cite{Giraud_1997,Giraud_1998}, who coupled the symmetric
viscous stress tensor to some new quantity, which evolved slowly in time, thus introducing
memory effects. This was simply achieved by adding two non-propagating distributions to
conventional LB models, such as the D2Q9 model (refer to Fig.~\ref{figure_3}), thus resulting in a modified D2Q11 model.
These two additional distributions, $f_9$ and $f_{10}$, which undertake a resting velocity, such that,
$\bm{\xi}_{9,10} = \bm{\xi}_{0}$, allow the capability to split the viscous stress tensor into two symmetric traceless
second-order tensors, for which different viscous memory effects can be incorporated. Predictions obtained using Chapman-Enskog analysis combined with numerical simulations of a pulsed Couette flow between two parallel plates demonstrated that the modified MRT approach obtained a linear Jeffrey's viscoelastic description \cite{Giraud_1998}.  This same model was subsequently coupled to a free energy model \cite{Orlandini_1995} by Wagner \textit{et al.} \cite{WAGNER2000227} and Wagner \cite{Wagner_2005} to simulate multiphase viscoelastic problems, specifically, a two-dimensional Newtonian bubble rising in a surrounding viscoelastic medium. This multiphase problem has emerged as a useful benchmark case owing to the well-documented experimental observations of a negative wake formation behind the Newtonian droplet, thus generating a cusped droplet shape \cite{Nature_Cusp,Liu_Liao_Joseph_1995}, as shown in Fig.~\ref{figure_5} c). Although the multiphase MRT model shows good qualitative agreement with experimental observations, a comprehensive quantitative validation remains lacking, primarily due to its two-dimensional numerical representation. Notably, Lallemand et al. \cite{Lallemand_2003} extended the MRT model to a full three-dimensional framework; however, their analysis remained largely theoretical.

A major limitation of LBM MRT models for viscoelastic flows is their lack of frame invariance, as shown through Chapman–Enskog analysis \cite{Phillips_2011}. In essence, these models are only valid in the linear viscoelastic regime and are therefore restricted to simple flow problems at low $Wi$ numbers. Additionally, they require increased memory to store extra distribution functions and involve multiple tunable control parameters, further complicating their implementation \cite{Chiyu_PRE,Lallemand_2003}.

\subsection{Linear Maxwell forcing models}
An alternative LBM approach to incorporate viscoelastic effects relies on a much simpler SRT collision operation coupled together with a linear Maxwell model forcing term. First formulated by Ispolatov \& Grant \cite{Ispolatov_2002}, this LBM approach considers the additional Maxwell-like (exponentially decaying) elastic forcing contribution,
\begin{equation}
    \bm{F}_{el}(\bm{x},t)=\bnabla\cdot\bm{\sigma}_p=\frac{\mu_p}{\lambda}\int_{-\infty}^{t} \exp\left(-\frac{t-t'}{\lambda}\right)\nabla^2\bm{u}(\bm{x},t')dt'.
\end{equation}
Notably, considering $\Delta t\ll\lambda$, a first-order Taylor expansion gives, $\exp\left(-\Delta t/\lambda\right)\approx 1 - \Delta t/\lambda$, which facilitates the time discretisation,
\begin{equation}
    \bm{F}_{el}(\bm{x},t+\Delta t)=\bm{F}_{el}(\bm{x},t)\left(1-\frac{\Delta t}{\lambda}\right)+\frac{\mu_p\Delta t}{\lambda}\nabla^2\bm{u}(\bm{x},t).
\end{equation}

The elastic forcing contribution $\bm{F}_{el}$ is easily coupled as an additional momentum contribution in LBM following the techniques discussed in Subsection~\ref{SS:Forces}. 

\begin{figure*}[t]
	\centering
	\includegraphics[width=0.99\textwidth]{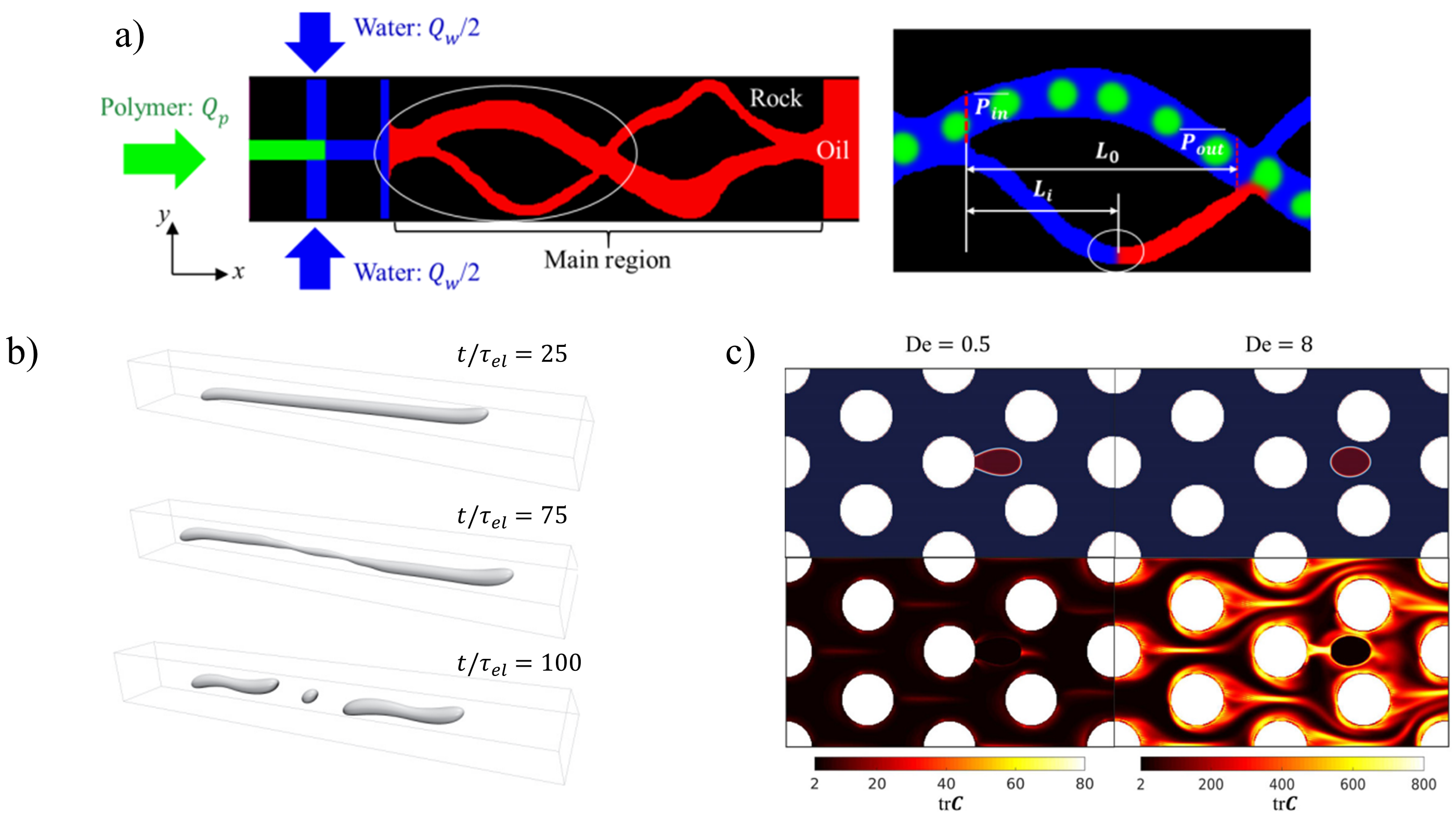}
	\caption{Applications of LBM to simulate complex and realistic multiphase viscoelastic flow interactions. a) Application of the linear Maxwell forcing model to investigate dispersed polymer flooding in complex pore channels (Image retrieved from Xie et al. \cite{Chiyu_PRE}). Here, a "polymer generation device" is created at the inlet to inject polymers (green), which are segmented into discontinuous drops by water streams (blue) from both sides. The addition of polymers helps recover initially saturated oil (red) within the realistic porous geometry. b) Hybrid LBM used to simulate the deformation and breakup of viscoelastic droplets in confined shear flow at $De=2$ (Image retrieved from Gupta et al. \cite{Gupta_Drop_PRE}). The additional polymer feedback forcing contribution plays a crucial role in stabilising droplet breakup. c) Hybrid LBM for simulating the mobilisation of a trapped oil droplet in porous media through viscoelasticity (Image retrieved from Dzanic et al. \cite{Dzanic_POF_Mobilization}). Sufficiently strong elastic effects induce polymer stresses near fluid–solid interfaces, contributing an additional forcing that drives a “pinch-off” mechanism to help displace trapped Newtonian droplets.}
	\label{figure_6}
\end{figure*}

In their work, Ispolatov and Grant \cite{Ispolatov_2002} numerically demonstrated the validity of their LBM approach for viscoelastic flow through a series of numerical simulations, including measurements of velocity autocorrelations, shear-wave propagation, and resonant shear oscillations, which were found to agree well with analytical predictions based on the Maxwell model. The model was subsequently extended by Yoshino \textit{et al.} \cite{YOSHINO} to two-phase flows within the framework of a free-energy lattice Boltzmann formulation. In this work, the extended model was applied to simulate the dynamics of a Newtonian droplet rising in, and being sheared by, a viscoelastic medium [refer to Fig.~\ref{figure_5} c) and d), respectively]. Building on this, Xie \textit{et al.} \cite{Chiyu_PRE} further generalised the approach to three-phase systems using a Rothman–Keller-type model for immiscible multiphase flows \cite{LECLAIRE2013318}. Moreover, through an extended MRT scheme, the authors were able to correctly describe the true momentum contribution for Maxwell materials, which involved setting the solvent viscous contribution $\mu_s=0$. In doing so, the model was successfully validated using the benchmark case of a Newtonian droplet rising in a viscoelastic medium for modest elastic effects \citep{Chiyu_PRE,xie_lei_balhoff_wang_chen_2021} [refer to Fig.~\ref{figure_5} c)]. Moreover, the model’s relative simplicity, along with its suitability for coupling to multiphase interactions, has facilitated its extension and application to complex flow problems involving realistic scenarios in porous media, such as dispersed polymer flooding in intricate pore networks \cite{Chiyu_PRE} [see Fig.~\ref{figure_6} a)], recovery of trapped non-wetting fluids through viscoelastic effects \cite{Chiyu_PHYS_REV}, and preferential flow control using dispersed polymers in heterogeneous porous media \cite{xie_lei_balhoff_wang_chen_2021}. Nevertheless, one notable limitation is the model’s assumption of linear elastic coupling, which may not fully capture the inherently nonlinear behaviours characteristic of viscoelastic systems \cite{Bird_Book}. In particular, under strong elastic effects (high $Wi$), the simple time derivatives employed in linear models are no longer frame invariant, limiting their ability to accurately describe strong polymer deformation \cite{Bird_Book,HINCH2021104668}.

\subsection{Lattice Fokker-Planck models}
An alternative view on incorporating the viscoelastic effects in LBM relies on a purely mesoscopic-based implementation  \cite{Onishi_2005, Onishi_2006}. The method adopts the fundamental discretisation approach of the LB method by deriving a discrete kinetic equation equivalent to the continuous Fokker-Planck equation. Similar to the discretised Boltzmann equation in Eq.~\ref{eq:BGK}, the Fokker-Planck equation involves describing the probability density of a given function, $\psi\left(\bm{r},\bm{x},t\right)$, which is used to describe the probability of finding a polymer dumbbell with configuration, $\bm{r}$, at a specific location, $\bm{x}$, in time, $t$. The discretised expression, to second-order accuracy using a discrete set of contributions, $\psi_j\left(\bm{r},\bm{x},t\right)$, with a configuration vector, ${r}_j$, can be seen below,
\begin{multline}\label{Laplace_Vel}
\psi_j\left(\bm{x},t+\Delta t\right)-\psi_j\left(\bm{x},t\right)=\\
-\frac{\Delta t}{\tau_{\psi}+0.5\Delta t}\left[\psi_j\left(\bm{x},t\right)-\psi_j^{\left(eq\right)}\left(\bm{x},t\right)\right]\\+\frac{\tau_{\psi}\Delta t}{\tau_{\psi}+0.5\Delta t}M_j\left(\bm{x},t\right)+\Delta\psi_j\left(\bm{x},t\right).
\end{multline}

Analogous to $\tau$ and $f_i^{eq}$ in Subsection~\ref{SS:LBM_Basics}, $\tau_\psi$ and $\psi_j^{\left(eq\right)}$ are the relaxation time and equilibrium distribution function, respectively. The final two terms account for solvent effects, whereby $M_j$ models the relative extension, contraction, and orientation of the dumbbells, with the full expression found in Onishi \textit{et al.} \cite{Onishi_2006}. Whereas, the final term related to the Laplace operator, $\Delta\psi_j$, describes the convection of the dumbbells \cite{Phillips_2011}. Macroscopic quantities, such as the polymer stress tensor, can be computed directly from the discrete distributions of the dumbbell configurations.
\begin{equation}
\bm{\sigma}_p=-n_p\sum_{j=0}r_jf_j^c\psi_j+n_p\sum_{j=0}r_jf_j^c\psi_j^{eq},
\end{equation}  
   
where $n_p$ is the number density of the dumbbells and $f_j^c$ is the spring force connecting the two beads, given by $f_j^c=hr_j$ with $h$ being a spring constant \cite{Onishi_2006}. Notably, the mesoscopic variables are easily redefined in terms of the relevant polymer quantities, namely, $\lambda=\tau_{\psi}$ and $\mu_p=(n_ph\tau_\psi)/c_s^2$. Different to traditional approaches, Onishi \textit{et al.} \cite{Onishi_2006} coupled the additional stress term directly into the equilibrium distribution function,
\begin{multline}\label{eq:Feq_Onishi}
f_{\alpha}^{eq} = w_{\alpha} \rho \biggl\{
1 + \frac{\bm{\xi}_{\alpha} \cdot \bm{u}^{eq}}{c_s^2}
+ \frac{1}{2} \left[
\left( \frac{\bm{\xi}_{\alpha} \cdot \bm{u}^{eq}}{c_s^2} \right)^2
- \frac{(u^{eq})^2}{c_s^2}
\right] \\
+ \left( \frac{\bm{\xi}_{\alpha} \cdot \bm{\xi}_{\alpha}}{2c_s^4}
- \frac{\bm{I}}{2c_s^2} \right) : \bm{\sigma}_p
\biggr\}
\end{multline}

Undertaking such a systematic coupling approach emanates the pressure coupling approach for nonideal contributions in multiphase problems by Swift \textit{et al.} \cite{Swift_1996}, whereby $f_{\alpha} ^{eq}$ leaves the zeroth and first-order moments unchanged, while modifying the second-order moment to include the additional polymer stress contribution, $\rho\bm{u}\bm{u}+\rho\bm{I}+\bm{\sigma}_p$ \cite{Onishi_2006, Gupta_Hybrid}. Through Chapman-Enskog expansion \citep{Onishi_2005}, it is shown that the scheme retains the Oldroyd-B polymer model (i.e., Eq.~\ref{eq:C_Eq} with $f=1$). However, more recently, Dzanic \textit{et al.} \cite{DZANIC_FEEDBACK} demonstrated that the pressure coupling approach in Eq.~\ref{eq:Feq_Onishi} leads to a macroscopic description with additional unphysical terms, which violate Galilean invariance, especially at high elastic effects.

Nevertheless, the Fokker-Planck model proposed by \cite{Onishi_2005} has shown good quantitative agreement with analytical solutions related to small amplitude oscillatory shear flow, as well as the four-roll mill benchmark problem \citep{OSMANLIC2016190}. Further model extensions, capable of simulating three-dimensional effects, were made by \cite{Onishi_2006}, which numerically simulated the deformation and orientation of viscoelastic drops. However, despite these advancements, in their work, \cite{Onishi_2006} acknowledged that the current formulation, which is based on a purely elastic Hookean dumbbell model, has limited applicability for strong shear flows. Recent extensions of this approach to the FENE-P model have been developed \cite{Fokker_FENEP}, however, given the general complexity of the Focker-Planck model, the additional inclusion of nonlinear polymer behaviour is far from trivial, and hence, relatively little work has been undertaken to apply this model in practical or large-scale viscoelastic flow simulations. Moreover, the idea of solving the polymer field through $\psi_j$ requires storing additional distribution functions related to a configuration vector $\boldsymbol{r}$, contributing to a memory-extensive process and a significant increase in computational cost.  

\subsection{Advection-diffusion schemes}
Following the development of the Fokker-Planck model, alternative approaches grounded in mesoscopic principles emerged, which involved undertaking a common advection-diffusion (AD) scheme typically applied as an extension to LB models for solving thermal problems \citep{Krueger}. The AD approach, initially applied by Denniston \textit{et al.} \cite{Denniston} and Marenduzzo \textit{et al.} \cite{Marenduzzo} to model the flow of liquid crystals, has since been adapted by Malaspinas \textit{et al.} \cite{MALASPINAS20101637} to numerically simulate viscoelastic flows. In principle, the LB arrangement can be used to solve for the polymeric stress term in viscoelastic fluids by simultaneously solving the hydrodynamic and conformation tensor quantities through LB discretisation \citep{MALASPINAS20101637,Phillips_2011}. More specifically, two separate LB equations are used to solve for the two properties, whereby the computational process involves computing each component ($\alpha,\beta$) of the rank-2 conformation tensor $\bm{C}$ by its own distribution function set $h_{i\alpha\beta}(\bm{x},t)$, using a modified AD formulation \citep{Phillips_2011}, 
\begin{equation}\label{eq:Fe_AD}
f_i \left(\bm{x}+\boldsymbol{\xi}_i \Delta t,t+\Delta t \right) = f_i(\bm{x},t) + \Omega_i(\bm{x},t)+{F}_i^{el}(\bm{x},t),
\end{equation}  
\begin{multline}\label{eq:ADE}
h_{i\alpha\beta} \left(\bm{x}+\boldsymbol{\xi}_i \Delta t,t+\Delta t \right) = h_{i\alpha\beta}(\bm{x},t) \\-\frac{\Delta t}{\tau_p} \left(h_{i\alpha\beta}(\bm{x},t) - h_{i\alpha\beta}^{eq}({{C}_{\alpha\beta}},\bm{u}) \right) \\
+ (1-\frac{\Delta t}{2\tau_p})\frac{{{G}_{\alpha\beta}}}{{{{C}_{\alpha\beta}}}}h_{i\alpha\beta}^{eq}({{C}_{\alpha\beta}},\bm{u}).
\end{multline}   
Here, a two-way coupling exists, whereby $\bm{\sigma}_p$ enters Eq.~\ref{eq:Fe_AD} through the forcing term $\bm{F}^{el}=\boldsymbol{\nabla}\bcdot\bm{\sigma}_p$, while the fluid velocity $\bm{u}$ helps evolve $h_{i\alpha\beta}$ in Eq.~\ref{eq:ADE}. Notably, ${G}_{\alpha\beta}$ is a source term depending on the polymer model, typically assocaited with the deformation and relaxation of polymers \cite{MALASPINAS20101637},
\begin{equation}
    \bm{G}=\bm{C}\bcdot\left(\bnabla\bm{u}\right)+\left(\bnabla\bm{u}\right)^{{T}} \bcdot \bm{C}-\frac{f}{\lambda}\left(\bm{C}-\bm{I}\right).
\end{equation}

The macroscopic quantity $\bm{C}$ can be computed through simple summation, ${C}_{\alpha\beta}=\sum\limits_{i=0}^{q-1} h_{i\alpha\beta}+{G}_{\alpha\beta}/2$, with a corresponding relaxation rate determined by the relaxation parameter, $\tau_p$. Analogous to the equilibrium distribution in the conventional LB model (Eq.~\ref{eq:Feq}), $h_{i\alpha\beta}^{eq}({{C}_{\alpha\beta}},\bm{u})$ represents the equilibrium distribution, which depends on the conformation tensor ${C}_{\alpha\beta}$. Considering the velocity field is externally imposed by the hydrodynamic LB solver $f_i$, the modified AD scheme ($h_{i\alpha\beta}$) does not require conserving the momentum field \citep{Krueger}. As a result, the equilibrium distribution is only truncated up to the $1$st-order term, leading to,
\begin{equation}\label{Heq}
h_{i\alpha\beta}^{eq}=w_i{C}_{\alpha\beta}\left(1+\frac{\boldsymbol{\xi}_i\cdot\boldsymbol{u}}{c_l^2}\right),
\end{equation}\\
where $c_l$, is the sound speed specific to the lattice structure used to solve the modified AD scheme. Notably, a lower order of truncation relaxes the requirements for rotational isotropy, thereby permitting the use of simpler lattice structures, such as D2Q5 \citep{Krueger}.   

Various attempts have been made to couple the discretised elastic force term ${F}_i^{el}$. For instance, Malaspinas \textit{et al.} \cite{MALASPINAS20101637} involved using the Guo \textit{et al.} \cite{Guo_2002} forcing scheme (Eq.~\ref{eq:Guo_Scheme} in Subsection~\ref{SS:Forces}).
Through Chapman-Enskog analysis, it is found that such an approach leads to a modified constitutive expression for the Oldroyd-B model,
\begin{multline}\label{Malas_CE}
\frac{\partial\bm{C}}{\partial t}+\bm{u}\bcdot\boldsymbol{\nabla} \bm{C}=\bm{C}\bcdot\left(\boldsymbol{\nabla}\bm{u}\right)+\bm{C}\bcdot\left(\boldsymbol{\nabla}\bm{u}\right)^T-\frac{1}{\lambda}\left(\bm{C}-\bm{I}\right)\\+\kappa\boldsymbol{\Delta}\bm{C}+\frac{\kappa}{c_l^2}\boldsymbol{\nabla}\bcdot\left(\bm{C}\frac{\partial\bm{u}}{\partial t}-\bm{u}\boldsymbol{\nabla}\bcdot\left(\bm{C}\bm{u}\right)\right).
\end{multline} 
The derived expression includes additional terms on the right-hand-side related to the polymer diffusion coefficient $\kappa$. Here, the Laplacian expression $\kappa\boldsymbol{\Delta}\bm{C}$ is directly related to an artificial diffusivity term typically included in viscoelastic solvers to enhance numerical stability \cite{Alves_Rev}. However, the final term, $\frac{\kappa}{c_l^2}\boldsymbol{\nabla}\bcdot\left(\bm{C}\frac{\partial\bm{u}}{\partial t}-\bm{u}\boldsymbol{\nabla}\bcdot\left(\bm{C}\boldsymbol{u}\right)\right)$, has no clear physical meaning and is an LB artefact \cite{MALASPINAS20101637}. This additional term was removed by Su \textit{et al.} \cite{SU201342} using a modified LB forcing scheme, which was first used by Shi \& Guo \cite{Shi_2009} to model nonlinear AD equations. 

In terms of application, the AD scheme has been numerically validated for a variety of different steady benchmark cases, including planar Poiseuille flow [Fig.~\ref{figure_5} a)] \cite{MALASPINAS20101637,SU_PRE,ZHANG2025106593}, 4:1 planar contraction \cite{SU_PRE,ZHANG2025105351}, viscoelastic flow
past a cylinder \cite{LEE201775}, lid-driven cavity \cite{SU201342,Ouallal_2025}, and four-roll mill problem [Fig.~\ref{figure_5} b)] \cite{MALASPINAS20101637,ZHANG2025106593}. Naturally, owing to the shared LB framework, extensions to multiphase problems are straightforward and have been demonstrated for a Newtonian droplet rising in a viscoelastic medium [Fig.~\ref{figure_5} c)], as well as a Newtonian (or viscoelastic) droplet suspended in a viscoelastic (or Newtonian) medium under shear flow [Fig.~\ref{figure_5} d)] \cite{Di_ADE_2019,Wang_2020_Shear_Bubble}. Notably, Ma \textit{et al.} \cite{Fang-Bao_2020} made further extensions to the AD scheme to include more complex fluid-structure interactions by coupling the immersed boundary-lattice Boltzmann method. However, these approaches have been mostly restricted to simulating moderate elastic effects (i.e., ${Wi}\sim1$) devoid of viscoelastic instabilities and for problems with relatively simple geometries. This is predominantly due to the AD scheme directly evolving the polymer field without careful numerical treatment to preserve the SPD properties of $\bm{C}$, a key requirement to overcome the well-known high Weissenberg number problem (HWNP) \cite{Alves_Rev}. Only very recently has this limitation been overcome through the LB flux-solver proposed by Zhang \textit{et al.} \cite{ZHANG2025105351}, which involves evolving 
polymer constitutive equations with a logarithmic representation, in line with conventional CFD numerical solvers \cite{Alves_Rev,VAITHIANATHAN20031,FATTAL2004281}. An alternative MRT-based stabilisation strategy was also recently proposed by Vaseghnia \textit{et al.} \cite{VASEGHNIA2025105467}, providing enhanced stability through the selective control of relaxation times and suppression of non-hydrodynamic modes. The authors successfully applied their model to simulate viscoelastic instabilities arising at higher $Wi$ in the four-roll mill problem. They note, however, that the optimal relaxation parameters in their MRT model are not universal and may require further tuning for different flow kinematics, higher Reynolds numbers, or transitional/turbulent viscoelastic regimes. Furthermore, the AD scheme shares a common constraint with the Fokker–Planck model: it requires storing and evolving an additional rank-2 tensor distribution function, leading to substantially increased memory demands, which can potentially be overcome with parallel scalability and careful memory management.

\begin{figure*}[t]
	\centering
	\includegraphics[width=0.8\textwidth]{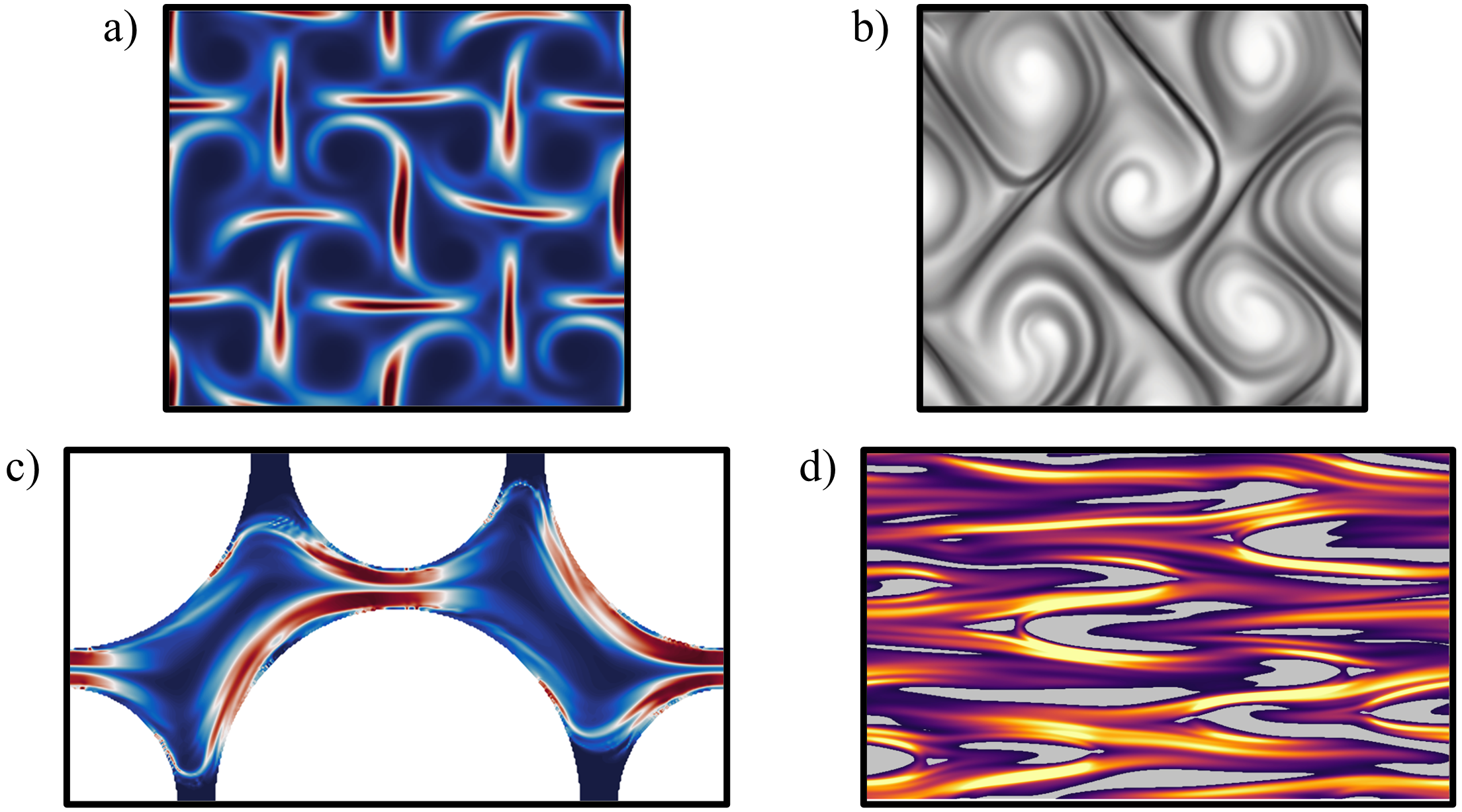}
	\caption{Examples of hybrid LBM applied to simulate challenging cases involving viscoelastic instabilities. Representative snapshots of the polymer field tr$\bm{C}$ for problems dealing with elastic turbulence, namely, a) four-roll mill (Image retrieved from Dzanic \textit{et al.} \cite{vedad_jfm}) and b) cellular forcing (Image retrieved from Dzanic \textit{et al.} \cite{Dzanic_PRE}). c) Polymer field tr$\bm{C}$ snapshot of inertialess viscoelastic instabilities arising in complex confined porous media geometries (Image retrieved from Dzanic \textit{et al.} \cite{Dzanic_PHYS_FLUID}). d) Elastoviscoplastic flows, which combine viscoelasticity and yield stress, in the elastic turbulence regime (Image retrieved from From \textit{et al.} \cite{From_Dzanic_Niasar_Sauret_2025}). Shown is a snapshot of tr$\bm{C}$ for the Kolmogorov forcing case in the turbulent regime, overlaid with the corresponding unyielded regions in grey.}
	\label{figure_7}
\end{figure*}
\subsection{Hybrid approaches}
Recently, a new class of viscoelastic LB solvers have emerged, which involve undertaking a hybrid numerical approach. In these schemes, the conventional LB method detailed in Section~\ref{sec:LBM} is used to simulate the hydrodynamic field, and is then coupled with an alternative conventional numerical scheme (i.e., finite difference schemes, finite volume methods, etc.), which resolves the polymer field. The idea of using two independent numerical schemes to simulate viscoelastic fluids becomes increasingly reasonable based on the previously outlined limitations regarding the pure mesoscopic LB formulations. First, the conventional numerical schemes used to resolve the polymer field are not required to share the same mesoscopic framework as the LB method, thus reducing any complexity regarding the coupling of the multiscale approach. Second, the computational cost associated with storing additional distribution functions related to $\bm{C}$ is not required for hybrid LB models, which directly solve for $\bm{C}$ in the polymer constitutive models. For instance, in the simplest hydrodynamic LB models, i.e., D2Q9 and D3Q15, mesoscopic viscoelastic LB approaches require storing $3\times9$ and $5\times15$ additional distribution functions associated with $\bm{C}$, respectively, whereas hybrid LB approaches store only the macroscopic information of $\bm{C}$. This feature significantly reduces the overall computational cost of these hybrid approaches \citep{SU_PRE}, which in turn, increases their potential applicability to solve more complex (and computationally expensive) viscoelastic problems. This same rationale has driven the switch from purely LB to hybrid approaches in other contexts, including liquid crystals and active matter \cite{Carenza2019,Marenduzzo}. Third, the application of conventional numerical schemes to resolve the polymer field enables the direct exploitation of well-established numerical stabilisation techniques originally developed for such solvers \cite{Alves_Rev}, thereby providing a straightforward route to overcome numerical challenges, such as the HWNP. In this way, hybrid LB approaches can leverage decades of progress made with conventional modelling tools, while retaining the intrinsic advantages of the LBM framework (see Subsection~\ref{sec:Why_LBM}).

Early attempts by Gupta \textit{et al.} \cite{Gupta_Hybrid} involved using an MRT D3Q19 LB model to resolve the hydrodynamic field, whereas a high-resolution finite difference scheme was used to simulate the polymer dynamics. Specifically, to enhance numerical stability at higher elastic effects (i.e., high $Wi$ numbers), the authors implemented the logarithmic Cholesky decomposition scheme \cite{VAITHIANATHAN20031} to preserve the SPD properties of $\bm{C}$, while further implementing an artificial diffusion term to regularise steep polymer stress gradients. Their model was validated across a range of rheological tests, including steady shear, elongational flows, transient shear, and oscillatory flows, and further demonstrated its capability to handle non-ideal multicomponent interfaces using the Shan–Chen interaction model (refer to Subsection~\ref{SS:Multiphase}). Notably, simulations of confined viscoelastic (Newtonian) droplets in Newtonian (viscoelastic) matrices under simple shear [Fig.~\ref{figure_5} d)] showed excellent agreement with theoretical predictions, highlighting the accuracy and robustness of the hybrid approach. The same model has also been applied to other microfluidic problems, including droplet formation and breakup in microfluidic junctions \cite{Gupta_Cross,Gupta2016} and confined shear flows \cite{Gupta_Drop_PRE} [refer to Fig.~\ref{figure_6} b)]. Dzanic \textit{et al.} \cite{DZANIC2022105280} further extended this hybrid approach by incorporating the high-resolution Kurganov–Tadmor scheme \cite{KT_SCHEME}, which improved the numerical stability in the presence of steep polymer stress gradients. This, in turn, allowed the authors to apply their model to simulate a variety of challenging high Wi number phenomena, such as elastic turbulence arising in the four-roll mill problem \cite{vedad_jfm} [Fig.~\ref{figure_7} a)], cellular forcing scheme \cite{Dzanic_PRE} [Fig.~\ref{figure_7} b)], and porous media geometries \cite{Dzanic_PHYS_FLUID} [Fig.~\ref{figure_7} c)]. The same model was also coupled with the Shan–Chen pseudopotential multicomponent model \cite{Dzanic_POF_Mobilization} (see Section~\ref{SS:Multiphase}) to simulate droplet mobilisation through porous media [Fig.~\ref{figure_6} c)]. Specifically,  the authors numerically demonstrated the existence of an additional momentum contribution sourced from the
viscoelastic fluid, which is capable of improving the mobilisation process of trapped oil droplets in porous media. More recently, the authors have extended their work to consider viscoelastic fluids which also exhibit a yield-stress behaviour, so-called elastoviscoplastic (EVP) fluids \cite{From_Dzanic_Niasar_Sauret_2025, Vedad_PNAS_Nexus, DzanicFrom2024}. By combining LBM with continuum-based models, such as the Saramito model \cite{SARAMITO20071}, the authors were able simulate EVP instabilities arising in elongational flow regimes, such as the four-roll mill and cellular forcing scheme \cite{DzanicFrom2024}, as well as investigate how plasticity modulates elastic instabilities in shear-driven flows \cite{Vedad_PNAS_Nexus}, such as Kolmogorov forcing [Fig.~\ref{figure_7} d)], even revealing a novel jamming phase transition \cite{From_Dzanic_Niasar_Sauret_2025}.

Alternative hybrid approaches have involved coupling LBM with other conventional CFD solvers, such as the finite volume method, to solve the constitutive equations for viscoelastic fluids. In their work, Zou \textit{et al.} \cite{ZOU201499} developed an integrated lattice
Boltzmann and finite volume technique for the simulation of isothermal and incompressible viscoelastic fluid flows (ILFVE). Specifically, a standardSRT LBM is adopted to solve the incompressible NSE, while a finite volume method with a second‑order accurate, cell‑centred discretisation and implicit time stepping is used to solve the polymer constitutive equations, including Oldroyd‑B and linear PTT models. Notably, ILFVE leverages widely used open‑source libraries, namely OpenLB for the LBM solver \cite{OpenLB} and OpenFOAM for the polymer solver \cite{OpenFoam}. The model was successfully validated for 2D Poiseuille flow and showed excellent numerical agreement with conventional numerical solvers when simulating the Taylor-Green vortex and 4:1 contraction flow at low to moderate elastic effects. Given that LBM inherently captures the pressure field, eliminating the need to solve an additional pressure-Poisson equation (see Subsection~\ref{sec:Why_LBM}), the authors demonstrated that ILFVE was significantly faster than comparable FVM PISO simulations, with ILFVE requiring approximately 20\% of the total PISO computation time.

More recently, Kuron \textit{et al.} \cite{Kuron} also developed a hybrid LBM–finite volume approach, specifically targeting Oldroyd-B fluids with complex and moving boundaries. In their framework, a TRT LBM is employed for the fluid flow, while the polymer stress evolution is solved using a finite volume discretisation on a collocated grid. Their model is validated for the time-dependent Poiseuille flow, steady shear flow, lid-driven cavity, and four-roll mill problem. Notably, the authors further extended upon previous viscoelastic LB models by including moving boundary conditions \cite{Ladd_1994}, enabling their model to simulate the sedimentation of a rotating sphere suspended in a viscoelastic medium.

\subsection{Challenges and future outlook}
Viscoelastic fluids represent a particularly complex class of non-Newtonian fluids, exhibiting complex rheological properties beyond simple shear-rate dependence. Their response includes phenomena such as normal stress differences, stress relaxation, and creep, which arise from the fluid’s elastic memory and microstructural dynamics. Capturing these effects numerically is challenging, as it requires resolving both the fluid’s viscous and elastic components over a range of time and length scales. LBM offers a natural framework for simulating such fluids due to its mesoscopic approach and flexibility in incorporating additional stress contributions. However, accurately reproducing viscoelastic behavior, especially under strong nonlinear deformations, remains computationally demanding and necessitates careful numerical treatment and validation. 

Early efforts to extend LBM toward fully mesoscopic viscoelastic formulations have followed several pathways. These include the use of modified MRT collision operators to include additional memory effects, lattice Fokker–Planck type models to evolve polymer configuration distributions directly on the lattice, and LB-based advection–diffusion schemes for transporting conformation tensors. Such approaches aim to retain the intrinsic strengths of LBM—locality, parallel scalability, and straightforward coupling between fluid and microstructural variables—while avoiding the need for external macroscopic solvers. Despite these promising developments, significant challenges remain. The aforementioned mesoscopic viscoelastic models frequently encounter numerical difficulties at frustratingly low Wi numbers due to their inability to conserve the SPD properties of the conformation tensor, thus leading to the growth of Hadamard instabilities that quickly overwhelm the calculation \cite{SURESHKUMAR199553}. Only very recently has this limitation begun to be addressed. Zhang \textit{et al.} \cite{ZHANG2025105351} proposed an LB flux-solver that evolves the polymer constitutive equations using a logarithmic stress formulation, aligning the method with established stabilisation strategies used in continuum CFD solvers \cite{Alves_Rev}. This represents a major step forward in improving the robustness of mesoscopic LB formulations at high elasticity. In parallel, an alternative MRT-based stabilisation approach was introduced by Vaseghnia \textit{et al.} \cite{VASEGHNIA2025105467}, which enhances stability through selective relaxation-time control and suppression of non-hydrodynamic modes. However, the authors noted that the optimal MRT relaxation parameters are flow-dependent and not universal, potentially requiring problem-specific tuning. In addition, both lattice–Fokker–Planck and advection–diffusion approaches remain computationally expensive due to the high dimensionality of the configuration space, incurring substantial memory costs as they require storing and evolving an additional rank-2 tensor distribution function. While such approaches may benefit from modern parallel architectures and advanced memory-management strategies, their scalability and efficiency relative to hybrid LBM–continuum solvers remains an open question.

The general constitutive polymer equation (Eq.~\ref{eq:C_Eq}) is inherently hyperbolic, lacking any numerical regularisation terms. The inclusion of an artificial polymer diffusion term $\kappa\boldsymbol{\Delta}\bm{C}$ to enhance numerical stability at high Wi numbers remains a topic of ongoing debate \cite{Alves_Rev}, as it can undesirably smear sharp stress gradients \cite{gupta_vincenzi_2019}, generate unphysical numerical artefacts \cite{vedad_jfm,Dzanic_PRE}, and laminarise viscoelastic instabilities \cite{gupta_vincenzi_2019,Dzanic_PRE,Dubief_REV}. While numerically challenging, conventional solvers can effectively define $\kappa=0$. By contrast, this flexibility is not available in mesoscopic LB formulations such as the advection–diffusion scheme, where a Chapman–Enskog expansion reveals that polymer diffusion arises intrinsically from the kinetic representation. In these models, $\kappa$ is directly linked to the relaxation time assigned to the conformation tensor distribution function $\kappa=c_s^2(\tau_p-\frac{1}{2})\frac{\Delta x^2}{\Delta t}$ and therefore cannot be eliminated, since $\tau_p>0.5$. As a result, such LB solvers cannot fully recover the hyperbolic nature of Eq.~\ref{eq:C_Eq}, limiting their applicability to resolve the important small-scale features of viscoelastic instabilities, such as elastic turbulence \cite{Steinberg_REVIEW} or elasto-inertial turbulence \cite{Dubief_REV}. Future work could explore the use of more advanced collision operators, such as MRT, to alleviate the intrinsic polymer diffusion, although these approaches would likely only mitigate, rather than eliminate, its effects.

Hybrid approaches that couple LBM with continuum solvers for the constitutive polymer dynamics offer a promising direction. Such formulations can, in principle, preserve the numerical stability and efficiency of the mesoscopic momentum solver while recovering the strictly hyperbolic character of the polymer equation. However, they introduce additional coupling complexity and may lose some of the self-contained elegance and fully localised, parallel nature of a purely mesoscopic LBM formulation. A systematic comparison between fully mesoscopic, hybrid, and purely macroscopic frameworks is still lacking; future studies should aim to quantify the numerical and physical implications of each approach to better assess their respective accuracies and computational trade-offs. A useful starting point would be a systematic comparison of different viscoelastic LB solvers for the simple benchmark case of flow past a cylinder in a channel, where notable discrepancies have been reported among continuum-based macroscopic models \cite{Alves_Rev}.

While the benefits of LBM for solving the momentum equation are well-established (see Subsection~\ref{sec:Why_LBM}), it is less clear whether the same advantages also extend to the polymer constitutive equations. Demonstrating the tangible benefits of a mesoscopic description of viscoelastic dynamics---whether through higher-order accuracy in representing polymer microstructural evolution or the ability to exploit fully localised operations on modern parallel hardware---is crucial to justify its continued development, particularly to the broader CFD community that often relies on well-established continuum-based solvers. Future work should aim to systematically benchmark and quantify these potential advantages, exploring scenarios where LBM offers unique value, such as high-resolution simulations of complex geometries, massively parallel computations on GPUs, or cases where hybrid and purely mesoscopic approaches may outperform traditional macroscopic frameworks. Even modest improvements in simplicity, scalability, or ease of coupling could make LBM an attractive alternative for challenging viscoelastic flow problems.

In terms of application, LBM models have proven particularly suitable for multicomponent systems where interfacial dynamics and phase separation are present, as they can capture essential features even with relatively simplified kinetic models. Significant practical progress involving viscoelastic fluids has been made in this direction, for example to investigate enhanced oil recovery through model porous media \cite{Chiyu_PRE,Dzanic_POF_Mobilization,xie_lei_balhoff_wang_chen_2021,Chiyu_PHYS_REV}, or to control droplet formation in flow-focusing microfluidic devices \cite{Gupta_Drop_PRE,Gupta_Cross, Gupta2016}. Future work on viscoelastic LBM should continue to exploit its inherent strengths, particularly for handling complex boundary conditions, such as those encountered in realistic porous media---scenarios where traditional hydrodynamic LBM has long proven highly effective \cite{BOEK20102305,Bakhshian2019, PRE_REALISTIC}. Furthermore, the relative ease of incorporating multiphysics within LBM makes it particularly appealing to couple viscoelastic LB models with heat transfer, mass transport, or electrokinetic effects. This enables the exploration of phenomena such as viscoelastic modulation of heat transfer or thermally driven flows in complex geometries, which remains an area of active research interest \cite{SASMAL_REV}. Moreover, given the recent surge in the application of LBM to active matter systems \cite{Carenza2019}, it becomes increasingly attractive to investigate complex problems at the intersection of activity and viscoelasticity. Examples include the collective dynamics and organisation of active constituents in viscoelastic environments \cite{Plan_Visc_Act}, such as tissue organisation within a viscoelastic extracellular matrix (ECM) \cite{EloseguiArtola2023}.

As a final remark, it is timely to continue to extend viscoelastic LB models toward more practical and industrially relevant applications. To date, the main obstacle has been the high computational cost associated with fully three-dimensional simulations of polymer constitutive equations, a limitation shared with conventional CFD approaches. To give a sense of the scale involved, state-of-the-art simulations of elastic turbulence in 3D channel flows using highly optimised spectral solvers \cite{Dedalus} have required on the order of $10^4$ CPU cores sustained for several hundred hours \cite{MOROZOV_PNAS}. Nevertheless, the intrinsically local and highly parallel structure of the LBM makes it particularly well suited to modern high-performance computing architectures, especially GPU-accelerated platforms \cite{latt2021palabos,LBM_GPU_LATT,latt2025multigpuaccelerationpalabosfluid}. As such hardware becomes increasingly accessible, these computational barriers are expected to diminish, opening the door to large-scale, three-dimensional simulations of viscoelastic flows in complex geometries.

\section{Conclusions}
\label{sec:Conclusions}
Non-Newtonian fluids play a ubiquitous role in nature, everyday life, and a variety of industrial and technological applications. Their complex rheological behaviour—including shear-thinning, shear-thickening, viscoplasticity, and viscoelasticity—poses significant challenges for experimental characterisation. Numerical modelling complements and aids experimental observations: continuum-based simulations provide expedient, computationally efficient predictions but are inherently limited to macroscopic scales, often missing finer microstructural details. Mesoscopic or microscopic frameworks, on the other hand, can capture these higher-order details and provide higher-fidelity descriptions of the underlying microstructural dynamics of non-Newtonian flows, but typically incur substantial computational cost. In this context, it is perhaps surprising that LBM can provide mesoscopic simulations with computational efficiency comparable to, and in some cases rivaling, continuum approaches, while still retaining the ability to resolve important microstructural interactions. These advantages, stemming from LBM’s inherent locality and natural parallelisability, combined with its flexibility in handling complex boundary conditions and including multiphysics, have led to its growing adoption for simulating non-Newtonian flows. 

In this review, we have offered an introduction to LBM for simple hydrodynamics, as well as its extensions to include additional non-Newtonian material responses. Specifically, in Section~\ref{sec:GNF}, we explored the application of LBM to generalized Newtonian fluids, encompassing shear-thinning, shear-thickening, and viscoplastic behaviours, highlighting both the numerical and practical application advancements. LBM has proven especially effective for simulating GNFs in complex geometries such as porous media, re-entrant corners, and vascular networks, as well as problems involving thermal transport, particle interactions, and fluid–structure interaction. However, several numerical challenges associated with capturing different GNF behaviour were highlighted, specifically the strong dependence of the collision relaxation time on local viscosity, numerical instabilities arising at very low or high shear rates, and the difficulty of maintaining accuracy and convergence for yield-stress fluids. For viscoelastic fluids in Section~\ref{sec:Viscoelastic}, we discussed various LB strategies used to evolve the microstructural evolution of polymers and capture elastic stresses. These approaches were primarily based on either fully mesoscopic methods inspired by traditional LBM, or coupled hybrid strategies that combine LBM with conventional continuum-based modeling frameworks. In comparing previous studies, fully mesoscopic LBM formulations often face numerical stability challenges and high computational costs due to the memory-intensive nature of evolving the polymer microstructure. Hybrid approaches seem to provide a practical compromise, retaining many of LBM’s advantages while improving numerical stability in resolving the polymer field. However, further work is still required to systematically compare these different LB approaches and quantify their relative benefits in terms of accuracy, stability, and computational efficiency.

Overall, LBM has continued to establish itself as a powerful CFD tool for simulating non-Newtonian flows, offering an attractive combination of computational efficiency and the ability to bridge macroscopic flow behaviour with underlying microstructural dynamics, while remaining adaptable to complex geometries and multiphysics. Continued methodological developments, together with systematic benchmarking against experimental and continuum-based results, will be essential to fully realise its potential and extend its applicability to increasingly complex rheological and industrially relevant systems.


\section*{CRediT authorship contribution statement}
\textbf{Vedad Dzanic:} Conceptualization, Writing – Original Draft, Review \& Editing, Visualization, Supervision, Project
administration. \textbf{Qiuxiang Huang:} Writing – Original Draft, Visualization, Review \& Editing. \textbf{Christopher Soriano From:} Writing – Original Draft, Review \& Editing. \textbf{Emilie Sauret:} Review \& Editing.

\section*{Declaration of competing interest}
The authors declare that they have no known competing financial interests or personal relationships that could have appeared to
influence the work reported in this paper. 

\section*{Acknowledgments}
The authors acknowledge the High-Performance Computing facilities at QUT. Prof. E. Sauret acknowledges support as an Australian Research Council Future Fellow (FT200100446). 

\section*{Data availability}
No data was used for the research described in the article.

 \bibliographystyle{elsarticle-num.bst}
 \bibliography{BIB_LBM}





\end{document}